\documentclass[english,reprint, prl, aps, superscriptaddress]{revtex4-1}
\usepackage[T1]{fontenc}
\usepackage[latin9]{inputenc}
\usepackage{color}
\usepackage{babel}
\usepackage{amsmath}
\usepackage{amssymb}
\usepackage{graphicx}
\usepackage{esint}
\usepackage[unicode=true,pdfusetitle,
 bookmarks=true,bookmarksnumbered=false,bookmarksopen=false,
 breaklinks=true,pdfborder={0 0 0},pdfborderstyle={},backref=false,colorlinks=true]
 {hyperref}
\hypersetup{
 pdfborderstyle={},citecolor=blue,urlcolor=blue,linkcolor=blue}



\begin{document}

\title{Photoionization and transient Wannier-Stark ladder in silicon: First principle simulations versus Keldysh theory}
\selectlanguage{english}
\author{Thibault J.-Y. Derrien}
\email{derrien@fzu.cz}
\affiliation{HiLASE Centre, Institute of Physics, Academy of Science of the Czech
Republic, Za Radnic\'{i} 828/5, 25241 Doln\'{i} B\v{r}e\v{z}any,
Czech Republic}

\author{Nicolas Tancogne-Dejean}

\affiliation{Max Planck Institute for the Structure and Dynamics of Matter (MPSD),
Luruper Chaussee 149, 22761 Hamburg, Germany}

\author{Vladimir P. Zhukov}

\affiliation{HiLASE Centre, Institute of Physics, Academy of Science of the Czech
Republic, Za Radnic\'{i} 828/5, 25241 Doln\'{i} B\v{r}e\v{z}any,
Czech Republic}
\affiliation{Federal Research Center for Information and Computational Technologies, 6 Lavrentyev Ave., 630090 Novosibirsk, Russia} 

\selectlanguage{english}%

\author{Heiko Appel}

\affiliation{Max Planck Institute for the Structure and Dynamics of Matter (MPSD),
Luruper Chaussee 149, 22761 Hamburg, Germany}

\author{Angel Rubio}

\affiliation{Max Planck Institute for the Structure and Dynamics of Matter (MPSD),
Luruper Chaussee 149, 22761 Hamburg, Germany}

\author{Nadezhda M. Bulgakova}
\email{bulgakova@fzu.cz}
\affiliation{HiLASE Centre, Institute of Physics, Academy of Science of the Czech
Republic, Za Radnic\'{i} 828/5, 25241 Doln\'{i} B\v{r}e\v{z}any,
Czech Republic}

\date{April 2021, 18th.}
\begin{abstract}
Nonlinear photoionization of dielectrics and semiconductors is widely treated in the frames of the Keldysh theory whose validity is limited to small photon energies compared to the band gap and relatively low laser intensities. The time-dependent density functional theory (TDDFT) simulations, which are free of these limitations, enable to gain insight into non-equilibrium dynamics of the electronic structure. Here we apply the TDDFT to investigate photoionization of silicon crystal by ultrashort laser pulses in a wide range of laser wavelengths and intensities and compare the results with predictions of the Keldysh theory.   Photoionization rates derived from the simulations considerably exceed the data obtained with the Keldysh theory within the validity range of the latter. Possible reasons of the discrepancy are discussed and we provide fundamental data on the photoionization rates beyond the limits of the Keldysh theory. By investigating the features of the Stark shift as a function of photon energy and laser field strength, a manifestation of the transient Wannier-Stark ladder states have been revealed which become blurred with increasing laser field strength. Finally, it is shown that the TDDFT simulations can potentially provide reliable data on the electron damping time that is of high importance for large-scale modeling. 
\end{abstract}
\maketitle

The Keldysh photoionization theory for atoms and solids published in 1964 \cite{Keldysh1964} plays a vital role in interpretation of ultrafast laser-induced phenomena, the field which continue to rapidly develop and influence our knowledge about fundamental processes in physics, chemistry, and biology. However, its applicability is limited by the requirements of 'sufficiently small' photon energy compared to ionization potential (band gap in solids) \cite{Keldysh1964,Vaidyanathan1979} and not too strong laser fields \cite{Deng2019b}.
For band-gap solids, the theory \cite{Keldysh1964} enables calculations
of the number of electrons excited from the valence to the conduction
bands per time unit, using a simplified description of the electronic levels reduced
to two bands. This simplification has become commonly employed for qualitative
simulations of laser-induced materials damage \cite{Kaiser2000,Chimier2011,Gallais2015} while experimental observations indicate that the Keldysh theory can considerably underestimate or overestimate photoionization rates, depending on irradiation regime and kind of solid \cite{Vaidyanathan1979,Lenzner1998,Gruzdev2014}. Direct experimental investigations of multiphoton inner excitation of band-gap materials was made possible relatively recently \cite{Popruzhenko2014,Otobe2008, Yabana2012a, Yamada2018}. Supported by comparisons with experimental data, several modifications of the Keldysh model for solids were
proposed to address discrepancies between experiment and theory \cite{Gruzdev2014,Shcheblanov2017a,Deng2019b,Otobe2019}.

An important feature of the original Keldysh theory is the dependence of the effective ionization potential for atoms and the effective band gap energy in the case of crystals, $U_{\text{eff}}$, on the laser field intensity \cite{Keldysh1964}. 
Although the increase
of the $U_{\text{eff}}$ value with intensity was proven experimentally
for atomic gases \cite{Pfeiffer2012,Anand2017}, an eventual increase
of the band gap energy in crystals \cite{Kaiser2000,Gulley2012a}
irradiated by linearly polarized light was, to our knowledge, not observed
for bulk materials. Recently, a clear shift of the energy levels
to higher values was found in monolayer WS$_2$ \cite{Sie2014} under irradiation by circularly polarized femtosecond laser pulses while a replication of the electronic levels by laser dressing was well demonstrated in GaAs
\cite{Schmidt2018}. Moreover, recent experimental studies performed for Si and ZnO \cite{Schultze2014,Winkler2017} supported a reduction of the
band gap energy in the laser field. 
These findings are consistent with the Keldysh theory where,  in the multiphotonic regime, $U_{\text{eff}}$ increases with the field amplitude ($\gamma \gg 1$ with $\gamma$ to be the Keldysh adiabadicity parameter) whereas in the tunneling regime the contribution of $U_{\text{eff}}$ fully fades out ($\gamma \ll 1 $). Nevertheless, the interpretations of $\gamma$ and $U_{\text{eff}}$ remain elusive, particularly for the cases of the mixed regime ($\gamma \sim 1$). 

Keldysh \cite{Keldysh1964} attributed the increase of $U_{\text{eff}}$ to the Stark effect. As a whole, the
Stark shift can transiently affect the band gap energy at high intensities while the laser dressing may eventually close the band gap, e.g., in regime when $\gamma \ll 1 $ \cite{Kwon2016}. However, there is no clarity of the manifestation and interplay of these effects at different irradiation regimes that is of high importance
for predicting the consequences of the action of ultrashort laser pulses on band-gap materials under the real experimental conditions \cite{Kaiser2000,Chimier2011,Gallais2015}. There is a need to verify the applicability of the
Keldysh theory to bulk materials in a wide range of the irradiation parameters and to find ways of determining the photoionization rates beyond its validity limits and beyond the perturbative regime \cite{Uemoto2019a}. 

In this Letter, 
we first confront the qualitative insights from the Keldysh theory with a Floquet model supported by the density functional theory (DFT). Then, we perform a quantitative comparison of the photoionization rates obtained in the frames of the Keldysh model and in the simulations based on first-principles time-dependent density functional theory (TDDFT) for a wide range of laser intensities and photon energies for silicon as an example.
Finally, based on the TDDFT simulations, we analyse the laser energy absorption and compare it with predictions of the Drude model. Means to improve accuracy of the Drude model for macroscopic description of materials optical response are discussed.

At the basis
of the Keldysh photoionization model, there is temporal averaging of a Hamiltonian which describes the interaction of the electromagnetic field with a two-level system,
representing an atom, a molecule or a solid in a simplified manner \cite{Keldysh1965a}.
When performing the temporal averaging of electronic energy levels over
one laser cycle \cite{Higuchi2014,DeGiovannini2016},
the so-called laser dressing manifests  as a replication
of the energy levels spaced by the photon energy. The discrete set of possible quasi-states appearing transiently in the energy gap  is usually referred as the Wannier-Stark ladder (WSL), the effect which can
be described qualitatively using a Floquet model \cite{FloquetModel}.

\begin{figure}
\begin{centering}
\includegraphics*[width=8.6cm]{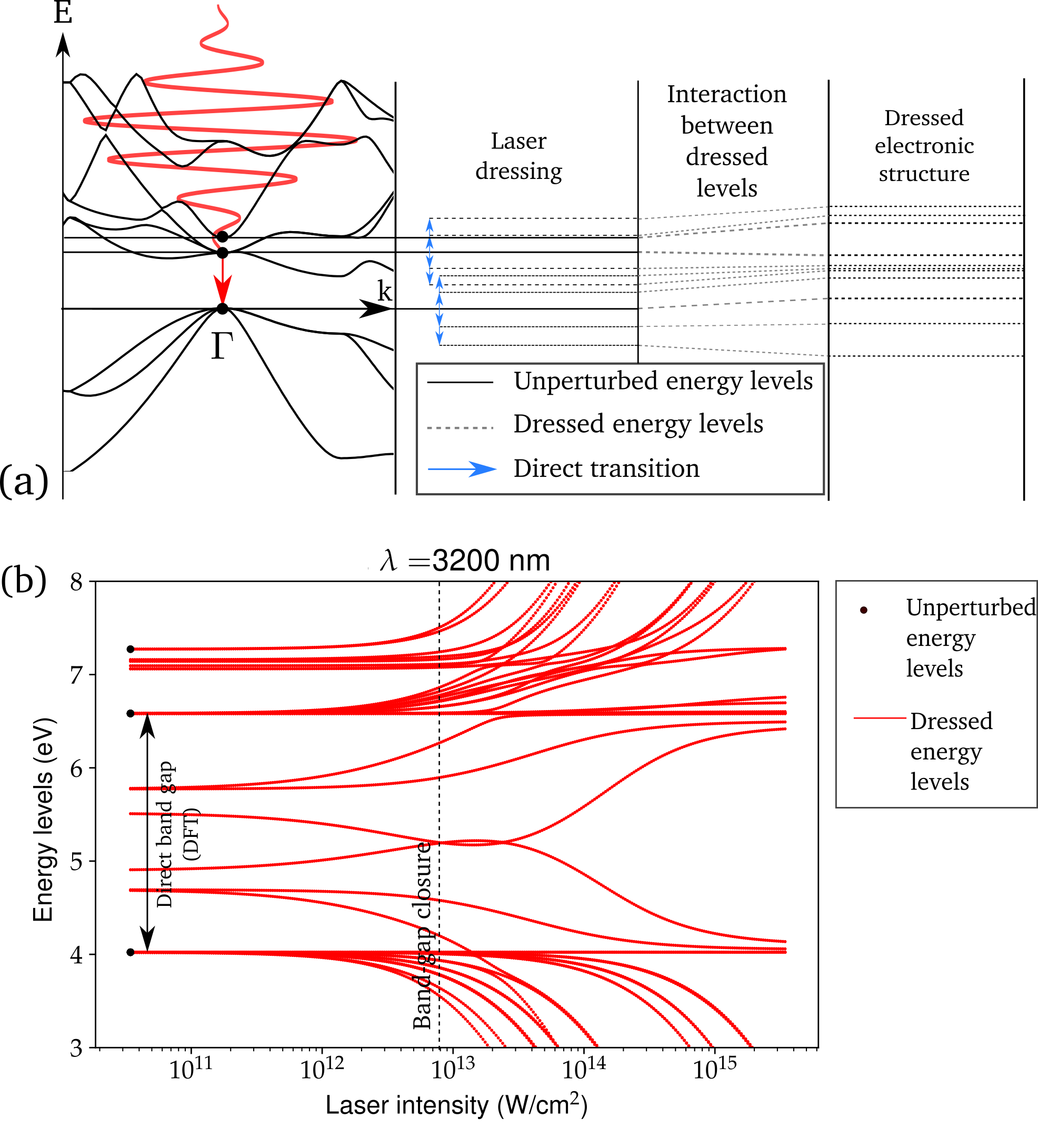}
\par\end{centering}
\caption{\label{fig:Stark-effect}Schematics (a) and realistic simulation (b) of a transient evolution of band structure in a band-gap solid on the example of a silicon crystal. The electronic bands ((a), left) are dressed by the laser field that results in replica bands at multiples of the driving photon energy (WSL, (a), right). With increasing laser intensity, a splitting of the electronic levels is observed and the valence and conduction bands experience anticrossing (b) at intensity close to $10^{13}$ W/cm$^2$ leading to transient metallization. See further details in the text.}
\end{figure}

Figure~\ref{fig:Stark-effect}(a) presents a schematics of the WSL
quasi-energy levels of silicon
shifted by the laser field under the combined action of laser dressing and the optical Stark effect.  The ground-state band structure shown on the left was calculated using density functional theory (DFT) (see Supplemental Material \cite{SupplMat}). Depending on the choice of laser polarization, wavelength and intensity, the distances between the quasi-energy levels
can either increase or decrease in presence of the strong optical
field as schematically shown on the right part of Fig.~~\ref{fig:Stark-effect}(a). Figure~\ref{fig:Stark-effect}(b) shows the calculated energies of the laser dressed states as a function of laser intensity, obtained using the Floquet model for the $m=8$ states at the $\Gamma$ point of Si. Note that the dipolar matrix elements \cite{FloquetModel} were computed using the DFT based on the local-density approximation (LDA). 
With increasing laser intensity,
the number of replicas, which are associated with the increasing contribution of the anharmonic motion of the electrons, also increases \cite{DeGiovannini2016}. With $n = 2$ replicas per ground electronic state, 
the Floquet model reveals multiple crossings (as well as anti-crossings, see \cite{Durach2011}) of the dressed electronic levels that may be manifested as a transient metallization during the laser pulse \cite{Durach2011}. The latter effect gives birth to important applications in ultrafast optoelectronics
\cite{Schiffrin2012,Schultze2012,Kwon2016}.

Let us next consider the results of numerical simulations for silicon irradiated by ultrashort laser pulses, which have been obtained by solving
the Kohn-Sham (KS) equations using the time-dependent density functional
theory (TDDFT) \cite{Andrade2015,Yabana2012a}. The laser field was introduced in the form of
a time-dependent vector potential (velocity gauge). The crystal Si <100> is modeled by the method of \emph{ab initio }pseudo-potentials \cite{Troullier1991}
(see details in Supplemental Material \cite{SupplMat}) involving periodic boundary
conditions. The band structure calculated based on the LDA yields the direct band gap for $\Gamma\rightarrow\Gamma$ transition to be equal to $E_{g}^{\text{\ensuremath{\Gamma}}}=2.56$ eV \cite{Waroquiers2013}. The approach enables to account directly for the effects
of the laser field on the valence and conduction electrons from first
principles, in particular, at intensities where a perturbative approach is not applicable \footnote{Applicability of perturbative nonlinear optics is believed to be valid  for relatively small optical driver fields as compared to a characteristic atomic field strength $E_{at}$ \cite{Boyd2008}. In Si,
the $E_{at}$ value can be estimated as $\sim$26 V/nm \cite{Bulgakova2005}. We note that, according to recent studies {\cite{Zheltikov2019}}, optical nonlinearities related to photoionization become nonperturbative at even smaller fields, which are defined by the Keldysh parameter $\gamma$.} and with accounting for a realistic band structure \cite{Lucchini2016,Otobe2016,Yamada2018}. 

Using TDDFT, the number of electrons excited from the valence
to the conduction bands, $n_{\text{exc}}$, was calculated using \cite{Otobe2008}
\begin{equation}
   n_{\text{exc}}(t)	=\frac{1}{V}\left[N_{\text{tot}}-\sum_{n,n',\boldsymbol{k}}\left|\int d^{3}\boldsymbol{r}\,\psi_{n',\boldsymbol{k}}^{^*}(\boldsymbol{r},t)\,\psi_{n,\boldsymbol{k}}^{\text{GS}}(\boldsymbol{r})\right|^{2}\right],
\end{equation}
where $n$ and $n'$ are band indices, $\boldsymbol{k}$ indexes the electron wave-vectors in 3 dimensions, $\psi_{n,\boldsymbol{k}}^{\text{GS}}(\boldsymbol{r})$ are the KS wave-functions of the ground state, and $\psi_{n',\boldsymbol{k}}^{^*}\left(\boldsymbol{r},t\right)$ are the KS time- and space-dependent wave-functions, where $^*$ indicates complex conjugation. $V$ is the volume of the simulation box (constant in this work),  $N_{\text{tot}}$ is the total number of electrons in the simulation volume, expressed by $N_{\text{tot}}=\sum_{n,\boldsymbol{k}}\left|\psi_{n,\boldsymbol{k}}^{\text{GS}}\left(\boldsymbol{r}\right)\right|^{2}$. The summing is performed over the occupied states in the valence bands. Note that in the ground state, before the laser pulse action, the obtained density of electrons in the conduction bands $n_{\text{exc}}\left(t=0\right)$ is zero. 
The gauge-related problems associated with the virtual excitations were avoided by exploiting the simulation results obtained shortly after the end of the laser irradiation  \cite{Otobe2008}.
The KS equation was solved numerically for a number of laser parameters by varying
the photon energy ${\hbar\omega}$ and the laser intensity $I$. Details of the
numerical approach are given in the Supplemental Material \cite{SupplMat}. 

\begin{figure*}
\begin{centering}
\includegraphics[width=12cm]{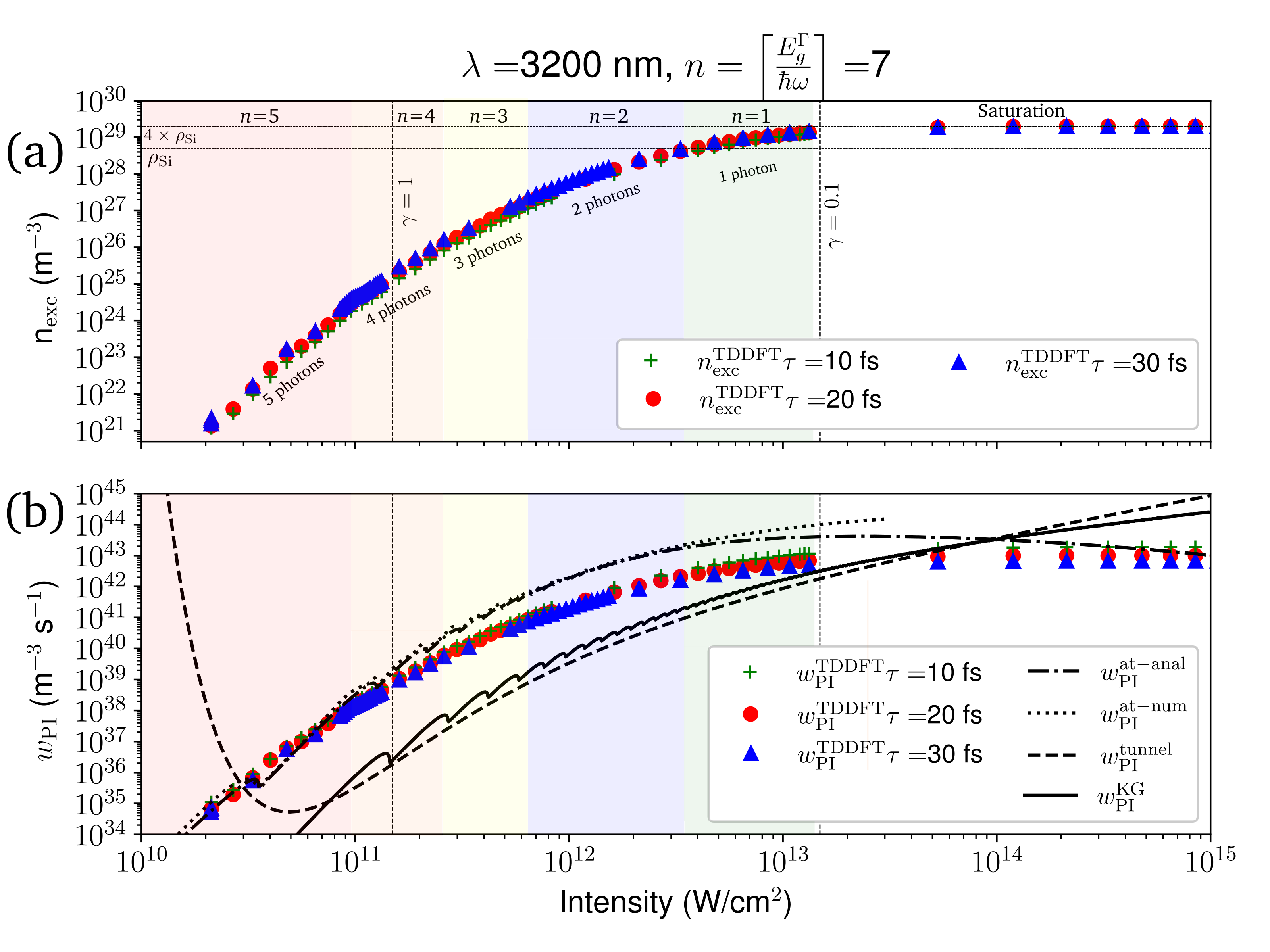} 
\par\end{centering}
\caption{\label{fig:TDDFT-3200nm}Density of electrons excited to the conduction bands $n_{\text{exc}}$ by the end of the laser pulse
(a) and the excitation rates $w_{\text{PI}}$ (b) as a function of the laser intensity in the bulk of silicon irradiated at $\lambda$ = 3200 nm. By dots, the results of the TDDFT simulations based on the LDA functional are presented for laser pulses of a top-hat temporal shape with durations of 10, 20 and 30 fs. In (b), the results of the
KG model ($w_{\text{PI}}^{\text{KG}}$, black solid line) and the tunneling limit ($w_{\text{PI}}^{\text{tunnel}}$, black dashed line) are added for comparison. Also for illustration, the data according to the analytical expression for the excitation rate for atoms \cite{Keldysh1964}, $w_{\text{PI}}^{\text{at-anal}}$, and a more exact numerical calculations of integrals of the Keldysh theory $w_{\text{PI}}^{\text{at-num}}$ are shown (see Supplemental Material \cite{SupplMat}). }
\end{figure*}

Figure \ref{fig:TDDFT-3200nm} presents the results of TDDFT simulations for 10, 20 and 30 fs laser pulse durations (a "softened" top-hat temporal shape, see details in the Supplemental Material \cite{SupplMat}) at the wavelength $\lambda$ = 3200 nm (${\hbar\omega}$ = 0.387 eV). The number of photons per electron for crossing the direct band gap can be estimated as $n \sim E_{g}^{\text{\ensuremath{\Gamma}}}/\hbar\omega$ that is 7 photons at this wavelength. Using the scaling law $n_{\text{exc}} \sim I^n$ \cite{Temnov2006}, the slope of the function $n_{\text{exc}}(I)$ in Fig. \ref{fig:TDDFT-3200nm}(a) can be associated with 5 photons that are
needed to cross the band gap at relatively low intensities in the range of $2\times 10^{10}-1.5\times 10^{11}$ W/cm$^2$ ($\gamma \gtrsim 1$) \footnote{For 1600 nm and 800 nm wavelengths (see \cite{SupplMat}), the slope of the $n_{\text{exc}}(I)$, obtained in the TDDFT simulations at low intensities, corresponds to $n= {\left\lceil E_{\text{g}}^{\Gamma}/\hbar\omega \right\rceil}$.}. With increasing the laser intensity, the simulations are consistent with experimental trends of a decrease of the $n_{\text{exc}}(I)$ slope \cite{Temnov2006}, which can be conventionally fitted with the $I^n$ scaling law at lower $n$ numbers. The corresponding intensity ranges with different $n$ are marked in Fig. \ref{fig:TDDFT-3200nm}(a) and shadowed by colors. For each $n$-range of intensity, an effective (pulse averaged) multiphoton ionization coefficient $\sigma_{n}$ can be estimated using a simplified formula
$\partial n_{\text{exc}}/\partial t \approx n_{\text{exc}}(\tau_p)/\tau_p =\sigma_{n}I^{n}/n\hbar\omega$ where $n_{\text{exc}}(\tau_p)$ is the number of the electrons excited to the conduction bands by the end of the laser pulse of duration $\tau_p$. For $\tau_p$ = 10 fs at $\lambda$ = 3200 nm, the TDDFT results can be fitted with $\sigma_{1}=2.93\times10^6$ m$^{-1}$, $\sigma_{2}=2\times10^{-10}$ m/W, $\sigma_{3}=5.25\times10^{-26}$ m$^3$W$^{-2}$, $\sigma_{4}=3.05\times10^{-41}$ m$^5$W$^{-3}$, and $\sigma_{5}=4.84\times10^{-56}$ m$^7$W$^{-4}$.

Note that in a purely multiphoton regime, the Keldysh theory predicts increasing $n$ with intensity (see Refs.
\cite{Keldysh1964,Kaiser2000,Gulley2012a}) while the total photoionization rate can be considered as a sum of $n$-photon processes with different $n$ \cite{Otobe2019}. Based on {\it ab initio} simulations, we interpret the reduction
of the photon number $n$ necessary for transition across the band gap as a mutual contribution
of laser dressing of electronic states and tunneling that reduces the {\it 
effective} band gap during the laser action. At high intensities when $\gamma < 0.1$, the quantity of electrons excited to the conduction bands saturates to the number of electrons available for ionizations (4 valence electrons per atom in Si). 

The corresponding excitation rates $w_{\mathrm{PI}}^{\mathrm{TDDFT}}$ are compared
with the Keldysh theory in Fig. \ref{fig:TDDFT-3200nm}(b). In both approaches, the effective electron mass was taken from the DFT calculations of Ref. 
\cite{LaflammeJanssen2016} ($m^* = 0.2226m_e$ with $m_e$ to be the electron mass in vacuum). It should be stressed that we use here the Keldysh theory for solids corrected by Gruzdev \cite{Gruzdev2014} and referred further as the KG model (black solid line in Fig. \ref{fig:TDDFT-3200nm}(b)). At laser intensities below 10$^{13}$ W/cm$^2$, the $w_{\mathrm{PI}}^{\mathrm{TDDFT}}$ values are considerably larger than the KG photoionization rates. This is in line with statements in \cite{Vaidyanathan1979,Gruzdev2014} where it was noticed that the Keldysh theory may underestimate the photoionization rate by order(s) of magnitude. At higher intensities, in the TDDFT simulations the saturation regime is achieved, determined by the number of valence electrons available for ionization. In this regime, the KG rate and its tunneling limit (dashed line in Fig. \ref{fig:TDDFT-3200nm}(b)  \footnote{The employed formula is available in Ref. \cite{Kaiser2000} and
in Supplemental Material \cite{SupplMat}}) approach and finally exceed $w_{\mathrm{PI}}^{\mathrm{TDDFT}}$.  

The results for the original atomic Keldysh theory, both its analytical expressions and numerical integration, are also added in Fig. \ref{fig:TDDFT-3200nm}(b) (dash-dotted and dotted lines respectively), obtained for a virtual atom with an effective ionization potential of 2.56 eV \cite{SupplMat}. Surprisingly, at intensities below 10$^{11}$ W/cm$^2$, the atomic theory fits reasonably well the TDDFT results although at higher intensities it deviates from the {\it ab initio} data. Important is that replacing the saddle-point approximation in the Keldysh
theory by an exact integration has a minor effect, contrary to a conjecture in \cite{Vaidyanathan1979}. As noted in \cite{Golin2014}, further advances of the Keldysh theory of photoinization for solids can be seen in introducing the realistic band gap structure. However, the rigorous TDDFT simulations presented here represent a solid alternative, both from fundamental and practical points of view. As an illustration, the TDDFT results and their comparisons with the Keldysh theory are given for $\lambda$ = 1600 nm and 800 nm in the Supplemental Material \cite{SupplMat}. 

Comparing $w_{\mathrm{PI}}^{\mathrm{KG}}(I)$ and $w_{\mathrm{PI}}^{\mathrm{TDDFT}}(I)$, one can notice that the KG model yields oscillations of the multiphoton excitation rate with $I$ while the TDDFT-calculated rates demonstrate rather smooth intensity dependences (Fig. \ref{fig:TDDFT-3200nm}(b)). This difference can also be explained by the two-band approximation applied in the KG model, while the TDDFT accounts for multiple bands. We note that the zeros of $n$th order Bessel function correspond to a regular suppression of the effective transition probability called dynamical localization or
destruction of tunneling \cite{Tamaya2019, Xia2020}. In a two-band model, this effect manifests itself as a swift decrease of the excitation rate at certain intensities, whereas in the case of a multiband description, the Bourget theorem secures that only a single transition may be simultaneously disabled. As a result, the functions $w_{\mathrm{PI}}(I)$ provided by the TDDFT multiband simulations are smoother than those obtained in the frames of the KG model. 

An important comment should be made on the excitation rates derived from the TDDFT simulations and calculated by the Keldysh theory for the saturation regime.  In the Keldysh approach, the ionization rate is calculated by averaging the probability over many laser cycles. In our \emph{ab-initio} numerical calculations, electron excitation follows the instantaneous laser field and, hence, it is strongly varying during the period of the electromagnetic wave. Furthermore, valence electrons, which are pushed out to the conduction bands at the first part of the laser pulse, can partially return back to the valence bands when the electric field of the wave changes direction \cite{Schiffrin2012,Sato2014}. Our estimations of the excitation rates given in Fig. \ref{fig:TDDFT-3200nm}(b) for three pulse durations correspond averaging over one, two, and three laser periods (note that one period for $\lambda$ = 3200 nm is $\sim 10.7$ fs). Thus, in the saturation regime when all or almost all valence electrons are excited to the conduction bands already during the first cycle, the $w_{\text{PI}}^{\text{TDDFT}}$ value at 10 fs looks to be more reliable although the excitation rates for all three pulse durations give the same result, transfer of all valence electrons to the conduction bands (Fig. \ref{fig:TDDFT-3200nm}(a)). Note that the KG model does not account for decreasing the number of valence electrons available for excitation into the conduction bands at high intensities and its direct use can lead to unphysically large number of electrons in the conduction bands (Fig. \ref{fig:TDDFT-3200nm}(b)). When a considerable fraction of valence electrons has been transferred to the conduction bands by the front part of the laser pulse, the excitation rate in the trailing edge of the pulse should naturally be decreasing by the dynamic factor $(1-n_\text{exc}V/N_\text{tot})$ (in our notations; see, e.g., \cite{Bulgakova2005}). Further still, the validity range of the KG theory is limited to single ionization per atom while the TDDFT simulations are not subjected to this limitation. 

Analysis of the TDDFT-calculated excitation rates as a function of the photon energy provides
insights into the WSL and the Stark shift in bulk crystals. 
\begin{figure}
\begin{centering}
\includegraphics[width=8.6cm]{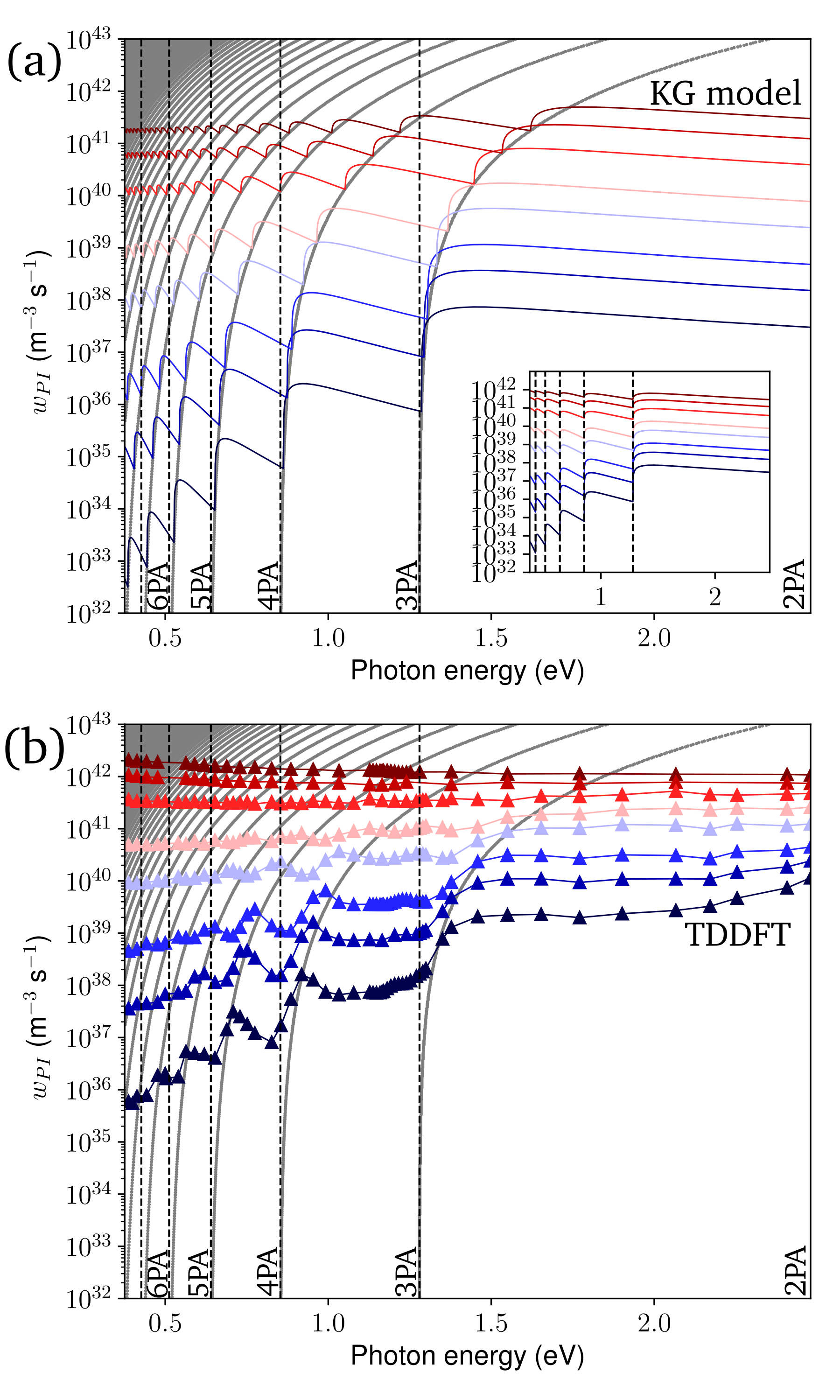}
\par\end{centering}
\caption{\label{fig:Excitation-rate-with-wavelength}Excitation rates $w_{\text{PI}}$
as a function of photon energy for $E$ = 0.5, 0.75, 1.0, 1.5, 2, 3, 4, 5 V/nm (curves from bottom to top), which correspond to $I$ = 0.03, 0.08, 0.13, 0.3, 0.53, 1.2, 2.1, 3.3 TW/cm$^2$, calculated using (a) the KG model
\cite{Gruzdev2014} and (b) the TDDFT (LDA functional, band gap of Si is 2.56 eV, pulse duration $\tau$ = 20 fs). $n$PA stands for ``$n$ photon absorption''
where $n$ is an integer.
The inset in (a) demonstrates the behaviour of the excitation rates when the Stark shift in the KG theory is disregarded.}
\end{figure}
Figure \ref{fig:Excitation-rate-with-wavelength} presents the comparison
of the excitation rates $w_{\text{PI}}$ calculated by the KG model (Fig. \ref{fig:Excitation-rate-with-wavelength}(a))
and by the TDDFT (Fig. \ref{fig:Excitation-rate-with-wavelength}(b))
for the $\Gamma\rightarrow\Gamma$ transition in Si. The data are given for the electric field amplitudes $E$ of the electromagnetic wave
from 0.5 V/nm to 5 V/nm (curves from bottom to top). The corresponding intensity range is 3.32$\times 10^{10} - 3.32\times 10^{12}$ W/cm$^2$ with the maximum intensity close to the saturation regime for $\lambda =$ 3200 nm (Fig. \ref{fig:TDDFT-3200nm}(a)).

At low electric fields, both the Keldysh formula and the TDDFT simulations reveal regular drops of the electron excitation at the photon energies close to resonances of the band gap at rest (marked by dashed vertical lines at $E_{g}^{\Gamma}/n$
with $n=\{2,3,4,5,6\}$ in Fig. \ref{fig:Excitation-rate-with-wavelength}). For each $n$ value in Fig. \ref{fig:Excitation-rate-with-wavelength}(a), the minima at different curves are connected by grey solid lines, indicating the Stark shift effect with increasing laser field strength (compare with inset in the figure where the Stark shift in the KG theory was disregarded). These grey lines are replicated to Fig. \ref{fig:Excitation-rate-with-wavelength}(b) to lead eyes, showing a similar Stark shift in TDDFT simulations. 
However, the TDDFT minima sometimes fall beyond the resonant photon energy (e.g. the minima seen between 3PA and 4PA in Fig. \ref{fig:Excitation-rate-with-wavelength}(b)).

Another qualitative difference between $w_{\text{PI}}^{\text{KG}}(\hbar\omega)$ and $w_{\text{PI}}^{\text{TDDFT}}(\hbar\omega)$ can be admitted for IR spectral range at high excitation fields.
In the TDDFT simulations the resonance drops are vanishing with increasing laser intensity (Fig. \ref{fig:Excitation-rate-with-wavelength}(b)). This is explained by a gradual transition
to the tunneling ionization regime. In addition, as noticed above, the KG model does not account for decreasing the number of valence electrons available for excitation into the conduction bands at high intensities that should contribute to vanishing the $w_{\text{PI}}$ resonances. We also note that coupling between moving
charges in the valence and conduction bands \cite{Higuchi2017,Schlaepfer2018}
can influence the excitation. The latter effect is not included
in the Keldysh theory but can be captured by other approaches \cite{McDonald2017}. 


\begin{figure}
\begin{centering}
\includegraphics[width=8.6cm]{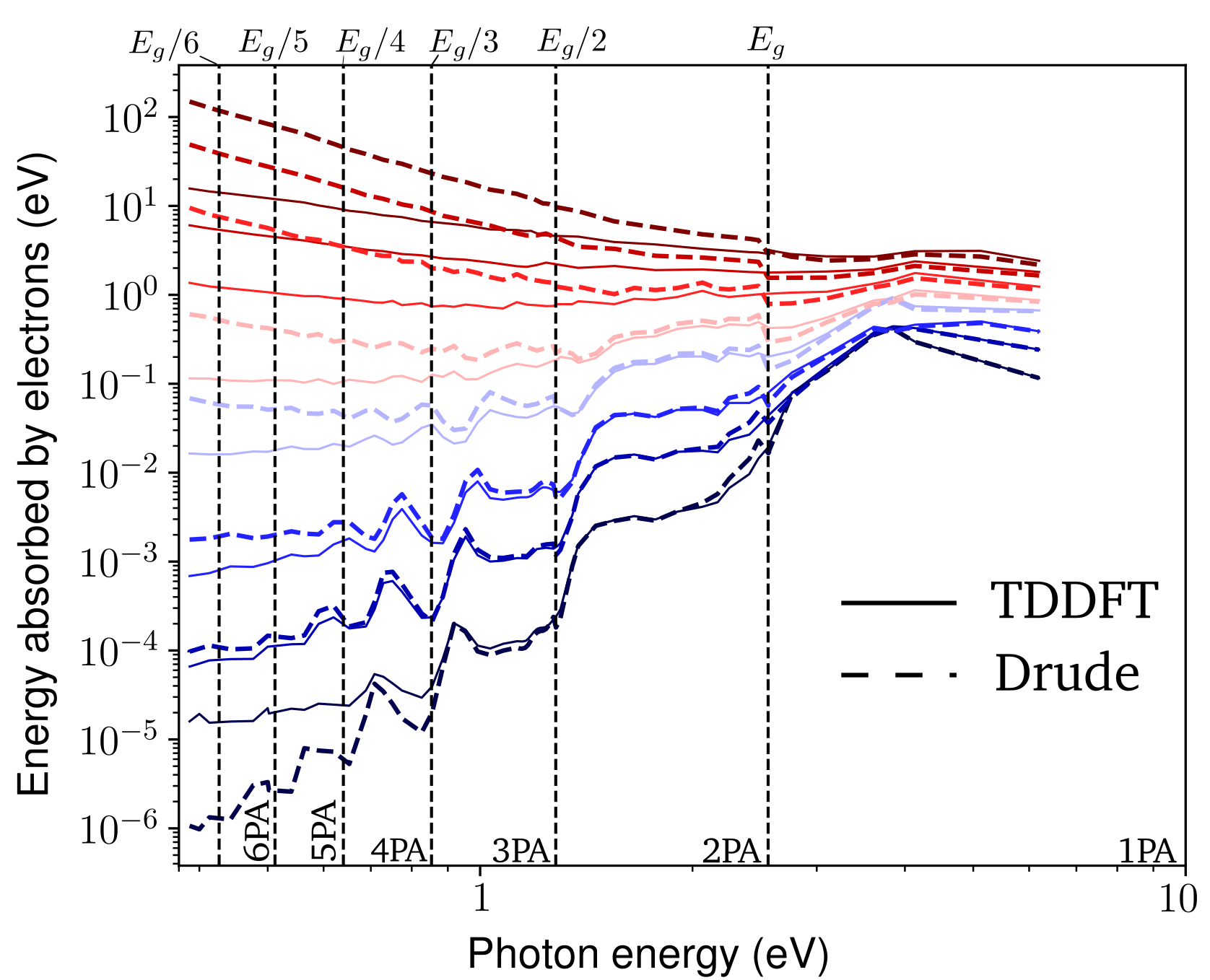}
\par\end{centering}
\caption{\label{fig:Absorbed-energy} Energy absorbed by electrons as a function
of photon energy for the laser field amplitudes of $E=$ 0.5,
0.75, 1.0, 1.5, 2, 3, 4, 5 V/nm (curves from bottom to top) obtained in the TDDFT simulations (solid lines),
and using a macroscopic Drude model for the densities of the excited electrons calculated by the TDDFT (dashed lines). The energy is given per one atom of silicon crystal. All calculations were performed for the top-hat temporal shape of laser pulses with duration of 20 fs.}
\end{figure}
In pump-probe and angle-resolved
photo-emission experiments \cite{Mahmood2016,Schmidt2018}, the possibility to observe the effects of quasi-energy levels (the WSL) in bulk band-gap crystals was recently proven. Here, based
on the TDDFT simulations of the laser energy absorbed by electrons (see solid lines in Fig.~\ref{fig:Absorbed-energy} and Supplemental Material \cite{SupplMat}),
we anticipate a possibility to directly observe the WSL by measuring the change in time-integrated optical transmission when accurately tuning the pump laser wavelength, preferably in the IR spectral range. Another fundamental aspect, which is of high importance for multiphysics large-scale modeling of the interaction of ultrashort laser pulses with band-gap materials, is an adequate description of dynamically changing reflectivity and absorptivity, which are usually treated in the frames of the Drude model \cite{Sokolowski-Tinten2000}. The TDDFT simulations give an excellent opportunity to verify on how accurate can be this simplest optical theory based on the classical equations of electron motion in an optical electric field.

The TDDFT and the Drude model have been compared based on calculations of the laser energy absorption at a wide range of laser parameters (Fig. \ref{fig:Absorbed-energy}). For this aim, the TDDFT-calculated density of the laser-excited electrons $n_{\text{exc}}^{\text{TDDFT}}$ was introduced into the Drude
formalism coupled with the energy balance equation for the electrons absorbing laser radiation. The latter was integrated for 20-fs laser pulses of the same shape as in the TDDFT simulations (see details in the Supplemental Material \cite{SupplMat}). By employing a Drude damping
time of $\tau_{D}$ = 6 fs for all tested wavelengths
and field strengths, it is possible to obtain a reasonable agreement with the TDDFT results (Fig. \ref{fig:Absorbed-energy}). A discrepancy between the Drude model and the TDDFT becomes pronounced for longer wavelengths and stronger optical fields. This can be attributed to the fact that $\tau_{D}$ is dependent on the density of electrons, their energy, and laser wavelength \cite{Sokolowski-Tinten2000,Sato2014a}. Thus, the TDDFT simulations can provide a possibility to derive the $\tau_{D}$ values for a wide range of laser irradiation conditions and for different materials. 

In conclusion, based on the TDDFT simulations we have investigated photoionization of crystalline silicon by ultrashort laser pulses in a wide range of laser intensities and for photon energies from UV to mid-IR spectral range. The excitation rates have been derived and compared with the Keldysh theory \cite{Keldysh1964}. The photoionization rates obtained within the the Keldysh approach in its validity range are smaller by more than order of magnitude as compared with the TDDFT simulation results that is in agreement with previous observations \cite{Vaidyanathan1979,Gruzdev2014} that can be attributed to several factors, including simplification of the band structure \cite{Golin2014}.   
We anticipate  that the TDDFT predicts well the photoionization process both within and beyond the limits of applicability of the Keldysh formalism. The excitation rates obtained in the first principle simulations represent valuable fundamental and practical information and, being tabulated in a wide range of the irradiation parameters, can be directly used in multiphysics large-scale
models, such as laser beam propagation in band-gap materials \cite{Sudrie2002,Bulgakova2013,Zhukov2017a}. 

By calculating the non-linear absorption as a function of photon energy for the Si crystal, we observed the quasi-energy levels known as the Wannier-Stark ladder. The energy of these laser-dressed states are shifted by the Stark effect at high optical field strengths. With further increasing field strength, these levels become less pronounced and finally disappear in the TDDFT simulation results. Finally, we have verified the Drude model based on laser energy absorption calculated in the frames of the TDDFT and a simplified energy balance equation, thus showing that the TDDFT simulations can potentially provide reliable data on the electron damping time as a function of density and energy of electrons and laser wavelength.  

It should be underlined that the results of the present work refer to pure photoionization-related processes in spatially homogeneous optical fields (dipolar approximation) which are not masked by electron-phonon coupling. For longer pulse durations than used here, the role of phonons becomes important, leading to indirect transitions, which are not accounted for in the present simulations. Furthermore, at longer timescales, lattice destabilization, non-thermal melting, Auger recombination, transport of hot quasi-free charge carries and electron energy relaxation are important for silicon and other semiconductors \cite{Shank1983a,Sokolowski-Tinten2000,Derrien2013,Sato2014a,Sato2014}, which are beyond the scope of this paper.

The research of T.J.-Y.D., V.P.Z. and N.M.B. is financed
by the European Regional Development Fund and the state budget of
the Czech Republic (project BIATRI: CZ.02.1.01/0.0/0.0/15\_003/0000445,
project HiLASE CoE: No. CZ.02.1.01/0.0/0.0/15\_006/0000674, programme NPU
I: project No. LO1602). T.J.-Y.D. also acknowledges funding from the European Commission for
the Marie Sklodowska-Curie Individual Fellowship, project No. 657424. This work was partially supported by the Ministry of Education, Youth and Sports from the Large Infrastructures for Research, Experimental Development and Innovations project "IT4Innovations National Supercomputing Center -- LM2015070". A.R. acknowledges financial support from
the European Research Council (ERC-2015-AdG-694097), Grupos Consolidados
(IT578-13), and European Union's H2020 program under GA no. 676580
(NOMAD). Access to storage facilities owned by parties and projects contributing to the National Grid Infrastructure MetaCentrum provided under the programme "Projects of Large Research, Development, and Innovations Infrastructures" (CESNET LM2015042), and to the DECI resource "Prometheus" based in Poland with support from the PRACE aisbl (project BOLERO) are greatly appreciated.

\bibliography{biblio}

\end{document}


\title{Photoionization and transient Wannier-Stark ladder in silicon: First principle simulations versus Keldysh theory. Supplemental Material}

\author{Thibault J.-Y. Derrien}
\email{derrien@fzu.cz}
\affiliation{HiLASE Centre, Institute of Physics, Academy of Science of the Czech
Republic, Za Radnic\'{i} 828/5, 25241 Doln\'{i} B\v{r}e\v{z}any,
Czech Republic}

\author{Nicolas Tancogne-Dejean}

\affiliation{Max Planck Institute for the Structure and Dynamics of Matter (MPSD),
Luruper Chaussee 149, 22761 Hamburg, Germany}

\author{Vladimir P. Zhukov}

\affiliation{HiLASE Centre, Institute of Physics, Academy of Science of the Czech
Republic, Za Radnic\'{i} 828/5, 25241 Doln\'{i} B\v{r}e\v{z}any,
Czech Republic}
\affiliation{Federal Research Center for Information and Computational Technologies, 6 Lavrentyev Ave., 630090 Novosibirsk, Russia} 

\selectlanguage{english}%

\author{Heiko Appel}

\affiliation{Max Planck Institute for the Structure and Dynamics of Matter (MPSD),
Luruper Chaussee 149, 22761 Hamburg, Germany}

\author{Angel Rubio}

\affiliation{Max Planck Institute for the Structure and Dynamics of Matter (MPSD),
Luruper Chaussee 149, 22761 Hamburg, Germany}

\author{Nadezhda M. Bulgakova}
\email{bulgakova@fzu.cz}
\affiliation{HiLASE Centre, Institute of Physics, Academy of Science of the Czech
Republic, Za Radnic\'{i} 828/5, 25241 Doln\'{i} B\v{r}e\v{z}any,
Czech Republic}

\maketitle

\section{Details of the numerical approach}

\subsection{Time-dependent Kohn-Sham equation}

The electrons in the crystal are modeled using the Kohn-Sham (KS)
auxiliary system that provides a one-to-one correspondence of the total electron density $n_{\mathrm{e}}\left(\boldsymbol{r},t\right)$
of the $N$ non-interacting electrons with the electron density of
the many-body problem \cite{Gross1996a,Marques2011}. We solve the time-dependent KS equations for the diamond structure of Si crystal along with applying periodic boundary
conditions in all directions. The dynamics of the $N$ electrons in
the crystal is described by solving a set of $N$ KS equations, expressed
by \cite{Ruggenthaler2018} 
\begin{gather}
\left[\left(-\frac{i\hbar}{2m_{e}}\boldsymbol{\boldsymbol{\nabla}_{r}}+\frac{\left|e\right|}{c}\boldsymbol{A}\left(t\right)\right)^{2}+\hat{v}_{\text{ion}}\left(\boldsymbol{r}\right)+\hat{v}_{\text{H}}\left[n_{\mathrm{e}}\left(\boldsymbol{r},t\right)\right]\left(\boldsymbol{r}\right)+\right.\label{eq:KSequation}\\
\left.+\hat{v}_{\text{xc}}\left[n_{\mathrm{e}}\left(\boldsymbol{r},t\right)\right]\left(\boldsymbol{r}\right)\right]\times\psi_{n,\boldsymbol{k}}\left(\boldsymbol{r},t\right)=i\hbar\frac{\partial}{\partial t}\psi_{n,\boldsymbol{k}}\left(\boldsymbol{r},t\right)\nonumber 
\end{gather}
where $\boldsymbol{\nabla_{r}}$ describes the real-space gradient
operator, $\hbar$ is the reduced Planck constant,
$m_{e}$ is the bare electron mass, $\left|e\right|$ is the elementary
charge, and $c$ is the light velocity in vacuum. Note that the non-local
contribution of the pseudo-potential to the external potential has
been omitted for simplicity. $\hat{v}_{\text{ion}}$ is the ionic
potential, $\hat{v}_{\text{H}}$ is the Hartree potential, $n_{\mathrm{e}}\left(\boldsymbol{r},t\right)$
is the \emph{total }density of electrons (i.e., all electrons in the valence and conduction bands) at the position $\boldsymbol{r}$
and instant $t$, and $\hat{v}_{\text{xc}}$ is the exchange-correlation
potential. $\psi_{n,\boldsymbol{k}}\left(\boldsymbol{r},t\right)$
is the time-dependent KS wave-function of an electron located in a
band $n$ with a wave-vector $\boldsymbol{k}$. The external vector
field $\boldsymbol{A}\left(t\right)$ is introduced in the velocity
gauge and is taken homogeneous in the simulation volume (dipolar approximation,
see Ref. \cite{Ullrich2012}). The ground-state KS wave-function
$\psi_{n,\boldsymbol{k}}^{\text{GS}}\left(\boldsymbol{r}\right)$
is obtained by solving self-consistently the static KS equations \cite{Ullrich2012}. The ground
state band structure is given by the energy representation of the
eigenvalues $\varepsilon_{n,\boldsymbol{k}}$ as a function of $\boldsymbol{k}$
taken along the path of high-symmetry points for Si crystal. 

\subsection{Calculations of the excited electron density\label{subsec:ExcitedElectronDensity}}

The excited electron density $n_{\text{exc}}\left(\boldsymbol{r},t\right)$
evolving in the conduction bands can be evaluated by a projection
of the occupied time-evolved KS orbitals $\psi_{n',\boldsymbol{k}}\left(\boldsymbol{r},t\right)$
on the ground state KS orbitals $\psi_{n,\boldsymbol{k}}^{\text{GS}}\left(\boldsymbol{r}\right)$
expressed by \citep{Otobe2008,Yabana2012}

\begin{align}
n_{\text{exc}}(t) & =\frac{1}{V}\left[N_{\text{tot}}-\sum_{n,n',\boldsymbol{k}}^{\mathrm{occ.}}\left|\int d^{3}\boldsymbol{r}\,\psi_{n',\boldsymbol{k}}^{^*}(\boldsymbol{r},t)\,\psi_{n,\boldsymbol{k}}^{\text{GS}}(\boldsymbol{r})\right|^{2}\right].\label{eq:ExcitedElDensityDefinition}
\end{align}
$V$ is the volume of the simulation box (constant in this work).
$N_{\text{tot}}$ is the total number of electrons in the simulation
box, expressed by $N_{\text{tot}}=\sum_{n,\boldsymbol{k}}\left|\psi_{n,\boldsymbol{k}}^{\text{GS}}\left(\boldsymbol{r}\right)\right|^{2}$.
Note that initially electrons are absent in the conduction bands. 

\subsection{TDDFT calculations of the excitation rates $w_{\text{PI}}$\label{subsec:Excitation-ratio}}

To avoid the gauge-dependence during the laser pulse, we use the results
obtained by the end of the laser pulse of duration $\tau_{p}$, and define a pulse-averaged
excitation rate $w_{\text{PI}}^{\text{TDDFT}}$ based on
the density of excited electrons, expressed via 
\begin{equation}
w_{\text{PI}}^{\text{TDDFT}}=\frac{n_{\text{exc}}\left(t=\tau_{p}\right)-n_{\text{exc}}\left(t=0\right)}{\tau_{p}}.\label{eq:AverageWpi}
\end{equation}
$n_{\text{exc}} \left(t=0\right) = 0$ in our case.

\subsection{Numerical details\label{subsec:Numerical-details}}

The presented computation results were obtained for a bulk Si sample.
We employed the primitive cell of Si composed of two atoms with an experimental lattice constant, $a_{0}=5.431$ \AA.
Non-orthogonal periodic boundary conditions in all directions were
used to describe the bulk crystal. Calculations of the ground state
Si were performed using a real-space discretization (with grid spacing
of 0.227 \AA) and converged using the local density
approximation (LDA) for the exchange-correlation functional. Several TDDFT calculations were performed using a more accurate but computationally demanding TB09 meta-generalized
gradient approximation for the functional \cite{Tran2009}. Assessment
of the validity of the LDA and TB09 functionals to describe the transient properties
of Si is provided in Refs. \citep{Gross1985,Godby1986,Sugino1995,Zuo1997,Waroquiers2013,Sato2015}. All calculations were carried out using the open source code Octopus \cite{Marques2003,Castro2006,Andrade2012}. The modeling of induced
fields was disabled. The atomic potential of Si is modeled using a norm-conserving
pseudo-potential \cite{Troullier1991}. 
The $\mathbf{k}$-grid was refined
until reaching a constant quantity of excited electrons. The convergence
was reached with using a $24\times24\times24$ grid in the $\mathbf{k}$-space. The integration of the KS equation was performed using the enforced time-reversal symmetry (ETRS) algorithm \cite{Castro2004}. 
The time step was reduced until convergence of the excited electron
density below 1\% variation achieved with a time step of 6.8 attoseconds for both functionals. Note that, upon using the TB09 functional, the temporal integration was performed based on predictor-corrector method \cite{Sato2015}. The TB09-based TDDFT results do not considerably differ from those obtained based on the LDA potential. However, further investigations are needed in this direction. 

\subsection{Introducing laser pulses with a top-hat temporal profile\label{sec:Description-of-top-hat}}

The electric field of the laser pulse is introduced in the KS equation using the dipolar
approximation \citep{Ullrich2012}. As a consequence, the time-dependent
laser field is homogeneous inside the simulation cell. The vector potential $\boldsymbol{A}\left(t\right)$ is linked with the electric
field $\boldsymbol{E}\left(t\right)$ via 
\begin{equation}
\boldsymbol{A}\left(t\right)=-c\int_{-\infty}^{t}\boldsymbol{E}\left(t'\right)\,dt'.
\end{equation}
If to approximate this integral using the slowly varying envelope approximation, the following
relation can be obtained $\boldsymbol{A}\left(t\right)\sim ic\boldsymbol{E}(t)/\omega.$

We are interested in the excitation rate of the electrons transferred
from the valence bands to the conduction bands. In order to provide this value on a similar basis as in the Keldysh theory where the ionization rate is averaged over laser cycles in a constant-amplitude field \cite{Keldysh1964}, we introduce a ``softened''
top-hat (STH) laser pulse of a duration $\tau_{p}$. The main part of STH pulse represents a plateau of the duration $\tau_{p}$ with rising and decay phases, both of the duration of $\tau_{r}$, which are short compared with  $\tau_{p}$.  The pulse is expressed
as: {\small{}
\begin{gather}
\boldsymbol{E}\left(t\right)=\boldsymbol{E}_{0}\cos\left(\omega t+\phi\right)\times\label{eq:PulseShape}\\
\times\begin{cases}
\sin^{2}\left[\frac{\pi\left(t+\tau_{p}/2+\tau_{r}\right)}{2\tau_{r}}\right], & \text{\ensuremath{-\frac{\tau_{p}}{2}-\tau_{r}\leq t\leq-\frac{\tau_{p}}{2}}}\\
\sin^{2}\left[\frac{\pi\left(t-\tau_{p}/2\right)}{2\tau_{r}}\right], & \frac{\tau_{p}}{2}\leq t\leq\frac{\tau_{p}}{2}+\tau_{r}\\
1, & \text{\ensuremath{\left|t\right|<\frac{\tau_{p}}{2}}}\\
0 & \text{\ensuremath{\left|t\right|\geq \frac{\tau_{p}}{2}}}+\tau_{r}.
\end{cases}
\end{gather}
}
The laser frequency is given by $\omega=2\pi c/\lambda$
and $\phi$ is the carrier envelope phase (CEP), which was set to 0. 

The rising time
$\tau_{r}$ for the STH pulses is used here to avoid a temporal
discontinuity in the field amplitude when the laser pulse starts and terminates. The $\tau_{r}$ value was optimized to obtain the minimum final excited
electron density. This optimum was reached for $\tau_{r}=4\pi/\omega$. 

\section{Details of the analytical models}

\subsection{Keldysh-Gruzdev excitation model for crystals\label{sec:Excitation-model}}

To calculate the number of excited electrons, several theories of
electron excitation for solids were proposed \citep{Mezel2010}. Gulley
\emph{et al}. summarized the Keldysh theory without amending modifications
\citep{Gulley2010,Gulley2012,Gulley2014,Gulley2014a}, whereas corrections
were proposed by Gruzdev \emph{et al.} \cite{Gruzdev2007,Gruzdev2010a,Gruzdev2014,Juergens2016},
to account for spin degeneracy. More recently, McDonald \emph{et al.}
\cite{McDonald2015,McDonald2017} have used a model based on semiconductor
Bloch equations \cite{Turkowski2008} to further study the effect
of band dispersion in electron excitation where not only the envelope
of the pulse but also the phase of the laser pulse can be accounted
for.

In this Section the analytical model is detailed, which was used to
calculate the $w_{\text{PI}}^{\text{KG}}$ values which are shown in Figs. 2--3 of the main manuscript. The Kane band structure, which is applicable for narrow band gap materials \citep{Gruzdev2014}, is considered in this model. 
In the general form, the nonlinear
photoionization rate $w_{\text{PI}}(I)$ in a constant-amplitude field is expressed as  
\begin{equation}
\frac{\partial n_{\text{exc}}(I)}{\partial t}=w_{\text{PI}}(I).\label{eq:ExcitationRate}
\end{equation}
According to the Keldysh-Gruzdev excitation model, the photoionization rates are expressed as 
\begin{gather}
w_{\text{PI}}^\text{KG}(I)=2\times\frac{2\omega}{9\pi}\left(\frac{m^*\omega}{\hbar F_{1}}\right)^{\frac{3}{2}}F_{0}\times\label{eq:AnalyticalKeldysh}\\
\exp\left(-\pi\left\lceil \frac{U_{\text{eff}}}{\hbar\omega}+1\right\rceil \times\frac{\hat{K}\left(F_{1}\right)-\hat{E}\left(F_{1}\right)}{\hat{E}\left(F_{2}\right)}\right)\nonumber 
\end{gather}
with 
\begin{gather}
F_{0}=\sqrt{\frac{\pi}{2\hat{K}\left(F_{2}\right)}}\sum_{n=0}^{\infty}\left[\exp\left(-\frac{\pi n\left[\hat{K}\left(F_{1}\right)-\hat{E}\left(F_{1}\right)\right]}{\hat{E}\left(F_{2}\right)}\right)\times\right.\label{eq:KeldyshGruzdevTerm}\\
\left.\times G\left(\pi\sqrt{\frac{\left\lceil \frac{U_{\text{eff}}}{\hbar\omega}+1\right\rceil -\frac{U_{\text{eff}}}{\hbar\omega}+n}{2\times\hat{K}\left(F_{2}\right)\times\hat{E}\left(F_{2}\right)}}\right)\right].\nonumber
\end{gather}
Here $m^* = 0.2226m_e$ is the electron effective mass with $m_e$ to be the electron mass in vacuum \cite{LaflammeJanssen2016}; $n = \lceil U_{\text{eff}}/\hbar\omega \rceil$ is the number of photons per electron for overcoming the effective potential barrier. Expression (\ref{eq:KeldyshGruzdevTerm}) employs the Dawson integral $G\left(z\right)=\int_{0}^{z}dy\,e^{y^{2}-z^{2}}$, which
is calculated numerically. $\hat{K}\left(x\right)$ and $\hat{E}\left(x\right)$
are the complete elliptic integrals of the first and second kind
respectively, which have the forms $\hat{K}\left(x\right)=\int_{0}^{\pi/2}\left[1-x^{2}\sin^{2}\theta\right]^{-1/2}d\theta$
and $\hat{E}\left(x\right)=\int_{0}^{\pi/2}\sqrt{1-x^{2}\sin^{2}\theta}d\theta$.
The effective ionization potential $U_{\text{eff}}$ accounts for
the energy shift induced by the Stark effect that can be written as $U_{\text{eff}}=\frac{2E_{g}}{\pi F_{1}}\hat{E}\left(F_{2}\right)$.
$F_1$ and $F_2$ are the functions of the adiabaticity parameter $\gamma$: $F_{1}=\gamma/ \sqrt{1+\gamma^{2}}$,
$F_{2}=F_1/\gamma$. The $\gamma$ value
is given by 
\begin{equation}
\gamma=\frac{\omega\sqrt{m^*E_{g}}}{eE_{\text{peak}}}=\frac{\omega\sqrt{m^*E_{g}}}{e\sqrt{\frac{2I_{\text{peak}}}{c\varepsilon_{0}}}}.\label{eq:AdiabadicityCoefficient}
\end{equation}
It characterizes irradiation 
regimes via the peak intensity $I_{\text{peak}}$ or the electric field amplitude $E_{\text{peak}}$.
When $\gamma\ll1$, the electron excitation from the valence band to the conduction band occurs via tunneling ionization
whereas $\gamma\gg1$ corresponds to multiphoton ionization. For $\gamma\ll1$, the Keldysh model for crystals describes
the tunneling mechanism of photoionization in the form 
\begin{gather}
w_{\text{PI}}^{\text{tun}}=\frac{2}{9\pi^{2}}\frac{E_{g}}{\hbar}\left(\frac{m^*E_{g}}{\hbar^{2}}\right)^{3/2}\left(\frac{e\hbar E_{\text{peak}}}{(m^*)^{1/2}E_{g}^{3/2}}\right)^{5/2}\label{eq:KeldyshTunneling}\\
\times\exp\left[-\frac{\pi}{2}\frac{(m^*)^{1/2}E_{g}^{3/2}}{e\hbar E_{\text{peak}}}\left(1-\frac{1}{8}\frac{m^*\omega^{2}E_{g}}{e^{2}E_{\text{peak}}^{2}}\right)\right],\nonumber 
\end{gather}
where $E_{g}$ is the bare band gap energy \cite{Gruzdev2014}. Note that this formula is not applicable at low fields where ionization is governed by the multiphoton mechanism \cite{Kaiser2000}. 

\subsection{Keldysh ionization model for atomic gases: analytical and numerical
integration \label{sec:Keldysh-excitation-model}}

In this section, the formulas of the Keldysh theory for ionization of an atomic gas are provided. We have employed the analytical atomic Keldysh model given by
Eqs. (16)-(18) from Ref. [\onlinecite{Keldysh1964}] to compare this theory with both the TDDFT and the Keldysh theory for excitation of band gap solids (see Fig. 2 of the main manuscript): 
\begin{gather}
w_{\text{PI}}^{\text{at}}=\rho\times\omega\sqrt{\frac{2E_{g}}{\hbar\omega}}\left(\frac{\gamma}{\sqrt{1+\gamma^{2}}}\right)^{3/2}S\left(\gamma,\frac{U_{\text{eff}}}{\hbar\omega}\right)\times\label{eq:AtomicKeldysh}\\
\times\exp\left[-\frac{2U_{\text{eff}}}{\hbar\omega}\frac{\text{arcsinh}\left(\gamma\right)-\gamma\sqrt{1+\gamma^{2}}}{1+2\gamma^{2}}\right],\nonumber 
\end{gather}
where 
\begin{gather*}
S\left(\gamma,x\right)=\sum_{n=0}^{\infty}\exp\left[\left(-2\left\lceil x+1\right\rceil -x+n\right)\times\right.\\
\left.\times\left(\text{arcsinh}\left(\gamma\right)-\frac{\gamma}{\sqrt{1+\gamma^{2}}}\right)\right]\times\\
\times G\left(\frac{2\gamma}{\sqrt{1+\gamma^{2}}}\sqrt{\left\lceil x+1\right\rceil -x+n}\right).
\end{gather*}
The Dawson integral $G\left(z\right)$ is given in Section \ref{sec:Excitation-model}.

\paragraph{Numerical integration.}

We use an estimative adaptation of material photoionization based on the
Keldysh theory for atoms \citep{Keldysh1964}. For this aim, we consider a virtual hydrogen-like atom with the
energy of its ground state equal to the material band gap, $E_{g}^{\text{\ensuremath{\Gamma}}}=2.56$ eV \cite{Waroquiers2013}. 
The electron of the atom has mass $m^*$, which is taken to be the same as in both the Keldysh theory for band gap solids (Eqs. (\ref{eq:ExcitationRate})-(\ref{eq:KeldyshGruzdevTerm})) and the TDDFT simulations. This
gives formulas for the ionization probability of atoms analogous to
Ref. [\onlinecite{Keldysh1964}]. Note that the ionization rate (Eq. (\ref{eq:AtomicKeldysh})) is multiplied by the atomic density of the solid $\rho$ (silicon in our case) to express the rate in m$^{-3}$s$^{-1}$. 

When solving Eq. (\ref{eq:AtomicKeldysh}), we calculated the integrals numerically instead of using the saddle-point method. 
The integrals have the following form
\[
P=\frac{\omega}{2\pi}\int_{0}^{2\pi/\omega}dt\,F\left(t\right)e^{i\eta\left(t\right)}.
\]
To calculate the integrals, a fine temporal grid $t_{k}=k\times\tau$ ($k=0,..., N_{t}$; $\tau=\frac{2\pi}{N_{t}\omega}$)
was used. Within the time segments $t_{k-1}<t<t_{k+1}$, the function $F$ is approximated
by the polynomial
\begin{gather*}
F\left(t\right)=F_{k}+\frac{F_{k+1}-F_{k-1}}{2\tau}\left(t-t_{k})+
\right.\\
\left.+\frac{F_{k+1}-2F_{k}+F_{k-1}}{2\tau^{2}}\left(t-t_{k}\right)^{2}\right.
\end{gather*}
For $\eta$, the linear expansion is used, $\eta=\eta_{k}+v_{k}\left(t-t_{k}\right)$
with $v_{k}=\frac{\eta_{k+1}-\eta_{k-1}}{2\tau}.$ Consequently, we have 
\begin{gather*}
P\simeq\frac{\omega}{2\pi}\sum_{k=1}^{N_{t}-1}e^{i\eta_{k}}\int_{-\tau}^{\tau}d\sigma e^{iv_{k}\sigma}\left[F_{k}+\frac{F_{k+1}-F_{k-1}}{2\tau}\sigma+\right.\\
\left.+\frac{F_{k+1}-2F_{k}+F_{k-1}}{2\tau^{2}}\sigma^{2}\right].
\end{gather*}
Such calculations of integrals are time-efficient and sufficiently precise, provided that the time step $\tau$ is much smaller than the laser cycle.

\section{The TDDFT results for different wavelengths; fitting of the KG model}

According to our TDDFT simulations, the first principles approach yields photoionization rates, which are at least an order of magnitude higher as compared to the Keldysh theory for solids. This has systematically been studied at different laser wavelengths, see Figs. 1--3. We remind that, in our study, the fields induced by the movement of charges in the time-propagation of KS orbitals were disregarded (Eq. (1)). Under this assumption, the laser pulse inside a bulk material is compressed  with the intensity multiplied by the material refractive index and it should be the same both in TDDFT simulations and in the Keldysh formulas. However, the TDDFT approach includes several important factors, which are absent in the Keldysh theory and thus can be responsible for the observed discrepancy.  

One of the most important factors is the realistic bandgap structure introduced in the TDDFT, which is dynamically varying upon electron excitation into the conduction bands with corresponding distortion of interatomic potential. As a result, the electron wave functions are subjected to the action of the field of the laser wave superimposed with the dynamic interatomic field. One can anticipate that the local field acting on the electronic component of the crystal is enhanced similarly to predictions of Gaier et al. \cite{Gaier2004} while not necessarily in a fluctuation manner. Another factor, also connected with the dynamic band structure, is the so-called laser
dressing of the electronic states  or, by other words, appearance of transient quasi-states usually referred as the Wannier-Stark ladder \cite{Schmidt2018}. We can also mention the Abraham-Minkowski problem \cite{Partanen2017} connected with the momentum of photons inside material, which still calls for investigations. Although this goes beyond the scope of the present study, a particular attention on the role of induced fields in the conservation of the momentum at interfaces is envisioned~\cite{Yabana2012}.

\begin{figure}
\begin{centering}
\includegraphics[width=8.6cm]{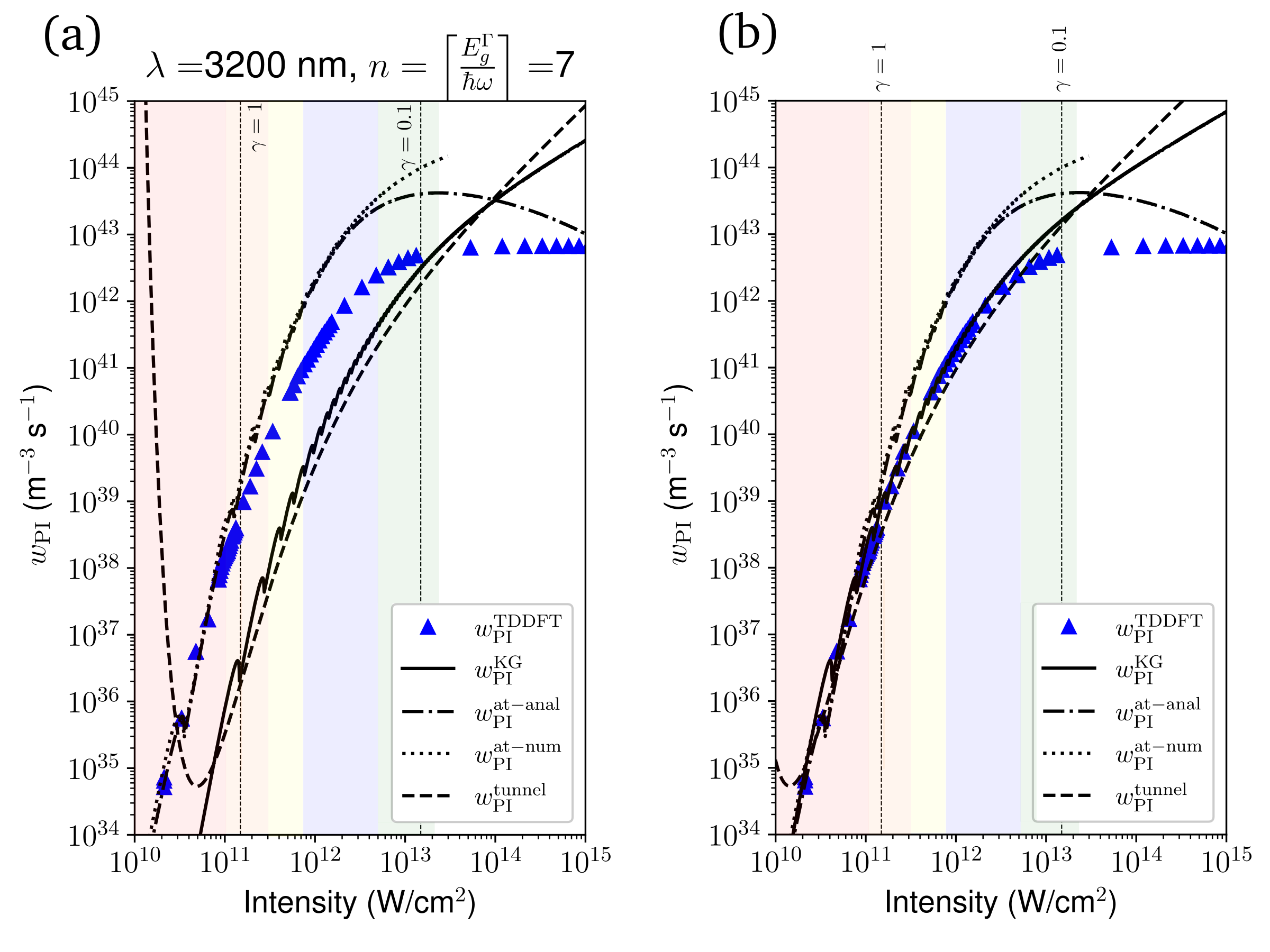} 
\par\end{centering}
\caption{\label{fig:3200-nm}(a) Comparison of photoionization rates as a function of laser intensity inside silicon crystal obtained in the TDDFT simulations, $w_{\text{PI}}^{\text{TDDFT}}$, after the laser pulse termination ($\tau_p$ = 30 fs, $\lambda$ = 3200 nm) with the analytical theories: the KG ionization rate $w_{\text{PI}}^{\text{KG}}$, the tunneling ionization rate, $w_{\text{PI}}^{\text{tunnel}}$, and the Keldysh photoionization rates for a virtual atom (Eq. (\ref{eq:AtomicKeldysh}), see text) obtained using the saddle-point method, $w_{\text{PI}}^{\text{at-anal}}$, and with numerically calculated integrals,  $w_{\text{PI}}^{\text{at-num}}$. (b) The same as in (a) but with fitting the KG and tunneling rates to the TDDFT results using normalization of the laser field in the KG formulas using the factor $\zeta = \sqrt{n(\lambda)}$ (see text).}
\end{figure}

\begin{figure*}

\begin{centering}
	\includegraphics[width=17.2cm]{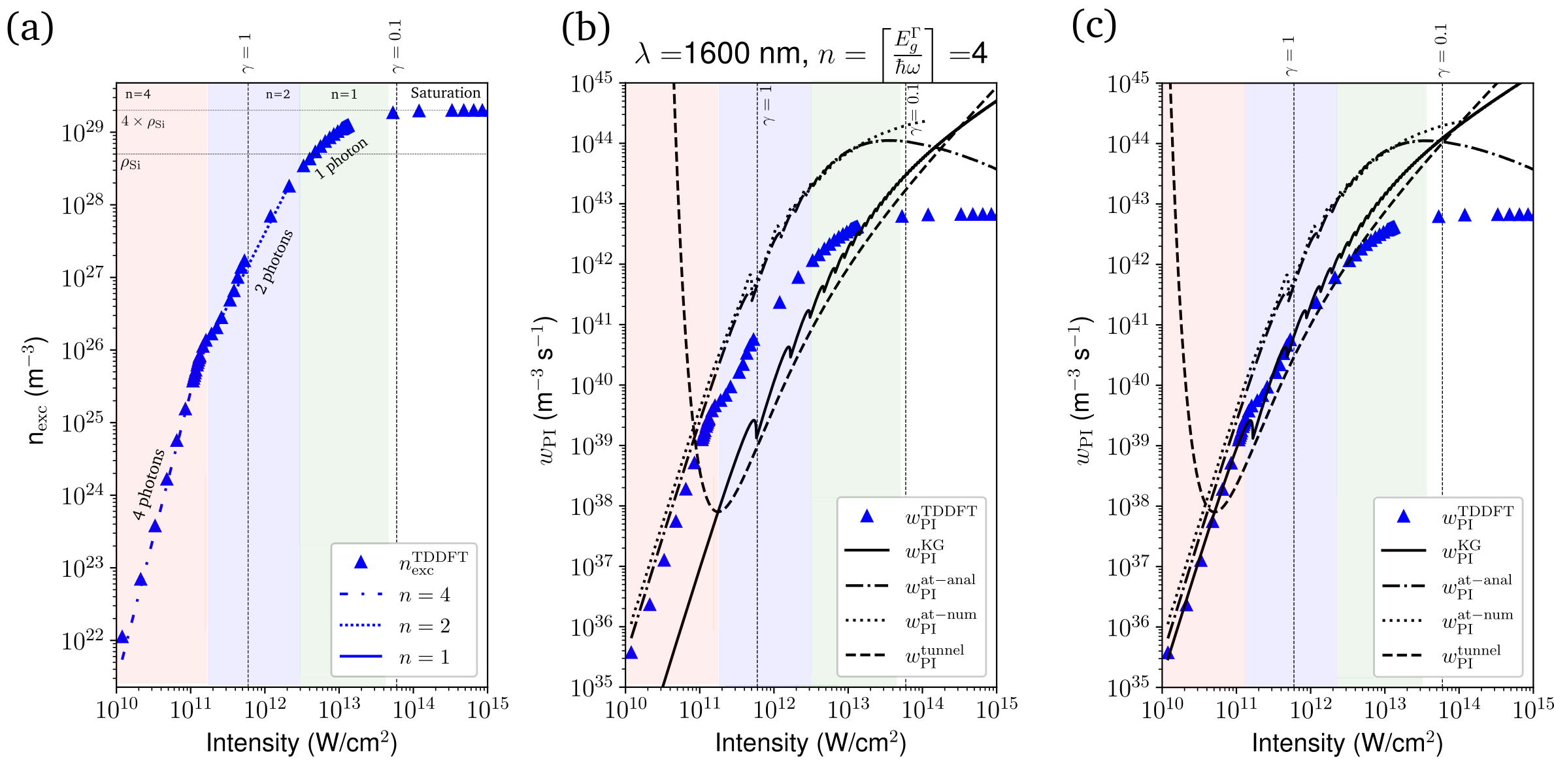} 
\par\end{centering}
\caption{\label{fig:Ibid-for-1600}The data for 1600 nm wavelength. (a) Electron density in the conduction bands after laser pulse termination ($\tau_p$ = 30 fs). Lines fitting the computed
$\sigma_{n}$ values are indicated. (b) and (c) are the same as in Figs. 1(a) and 1(b) respectively for another wavelength. }
\end{figure*}

\begin{figure*}
\begin{centering}
	\includegraphics[width=17.2cm]{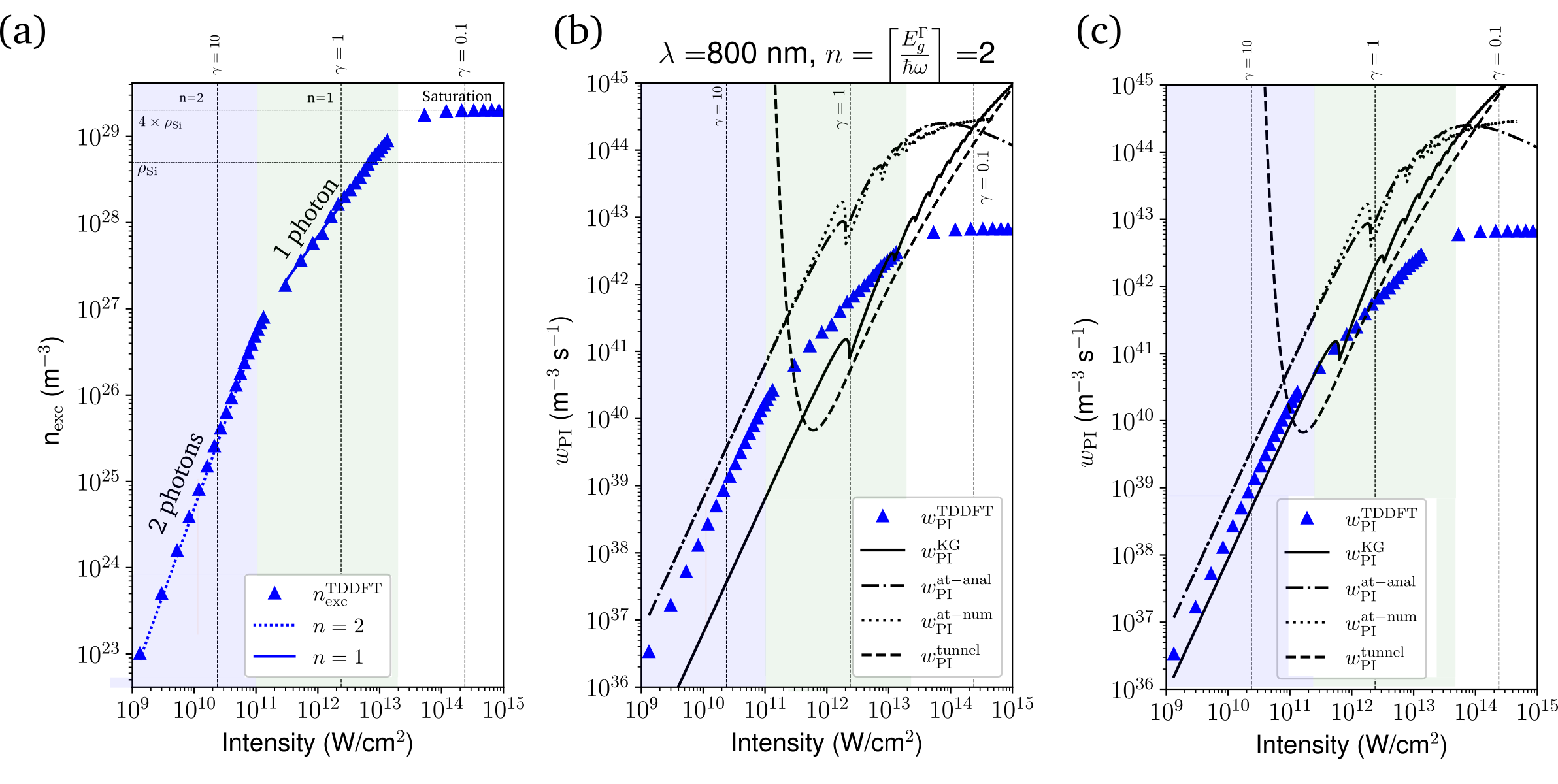} 
\par\end{centering}
\caption{\label{fig:Ibid-for-800}The same as in Fig. \ref{fig:Ibid-for-1600} for 800 nm wavelength.}
\end{figure*}

Figures 2(b) and 3(b) show that the KG model underestimates the results of the TDDFT simulations where the band structure is more realistic and laser dressing is naturally addressed. Interesting is to find a factor $\zeta$ of "laser field amplification", $E \rightarrow \zeta E$, at which the KG model would fit the TDDFT results. We have performed this procedure for different wavelengths and the results are presented in Figs. 1(b), 2(c), and 3(c) respectively for 3200, 1600, and 800 nm for both the KG model and its tunneling limit. By choosing the $\zeta$ value, it was surprisingly found that the accurate enough fits were achieved at $\zeta = \sqrt{n(\lambda)}$ where $n$ is the refractive index at the corresponding wavelength $\lambda$. It is not clear yet if it is a pure coincidence or it hides a physical cause. Also interesting is that for all studied wavelengths the atomic Keldysh theory applied for our virtual atom agrees reasonably with the TDDFT simulation results at low intensities (in the multiphoton regime, see Figs. \ref{fig:3200-nm}--\ref{fig:Ibid-for-800}). 

\section{Laser dressing, effect of Rabi oscillations and the role of Bessel
functions in strong field excitation}

During laser action, response of a bandgap material to irradiation can be described by shifting the energy levels of the band
structure as compared with the ground state structure \citep{Fritsche1966,DeGiovannini2016}. This shift is known as laser dressing and it originates from the coupling
of laser light to the electrons, which lasts the time of the laser illumination.
As a result of the light-induced periodic electron motion, the dressed
band structure does not represent anymore purely electronic states.
Instead, the dressed energy levels are considered as quasi-particles named
polaritons. This vision is not new and it was already employed in a number
of formalisms, e.g., already in the Keldysh theory \citep{Keldysh1964},
in atomic physics \citep{Cohen-Tannoudji1977a} and in the non-equilibrium
solid state physics \citep{Flick2019c,Rokaj}. Laser dressing may lead to ultrafast transient metallization \citep{Durach2010,Durach2011}, a phenomenon that has enabled
the development of novel applications in ultrafast optoelectronics
\citep{Kwon2016}. 

As stated in \cite{Durach2010}, a qualitative description of the laser dressing effect is possible from the knowledge of
the electronic ground state. We underline that the approach employed here is based on a simplified model Hamiltonian. In particular,
the importance of the term scaling in $A^{2}$ in the employed Hamiltonian, which is
disregarded in our approach, was investigated in a series of recent publications
\citep{Rokaj2018,Rokaj}. Also the dipolar matrix elements should be affected by intense laser fields,
a phenomenon that is not described in the simplified method we have employed to prepare Fig. 1 of the main manuscript, which therefore should be considered as a contextual illustration. 

In Fig. 3(a,b) of the main manuscript, the minima of the excitation rates $w_{\text{PI}}^{KG}(\hbar\omega)$ provided by the KG theory are outlined by grey lines. 
These reductions of $w_{\text{PI}}(\hbar\omega)$ are pronounced in the Keldysh theory and are also visible in the TDDFT simulation results, though mildly. In literature,
the origin of such reductions of the amplitude probability $w_{\text{PI}}(\hbar\omega)$
was attributed to the suppression of tunneling \citep{Tamaya2019,Kato2020,Xia2020}
(equivalently, these are described in the real-space as Wannier-Stark localization
\citep{Wannier1960,Wannier1969,Schmidt2018}). The tunneling drop
interpretation can be illustrated by an analytical solution constructed
from a simplified Hamiltonian model describing the effect of the laser
field on the conduction electronic subsystem \citep{Tamaya2019,Kato2020,Xia2020}. Using a rigid band viewpoint, a series of the Bessel functions of $n$-th
order appear in the corresponding analytical solution as a multiplicative term
for each replication order $n$ \citep{Tamaya2019,Kato2020,Xia2020}.
Since the Bessel function is inside the sum of possible wave-functions,
an eventual suppression of interband transitions depends on the number
of available electronic levels. 

\begin{figure}
\begin{centering}
\includegraphics[width=8.6cm]{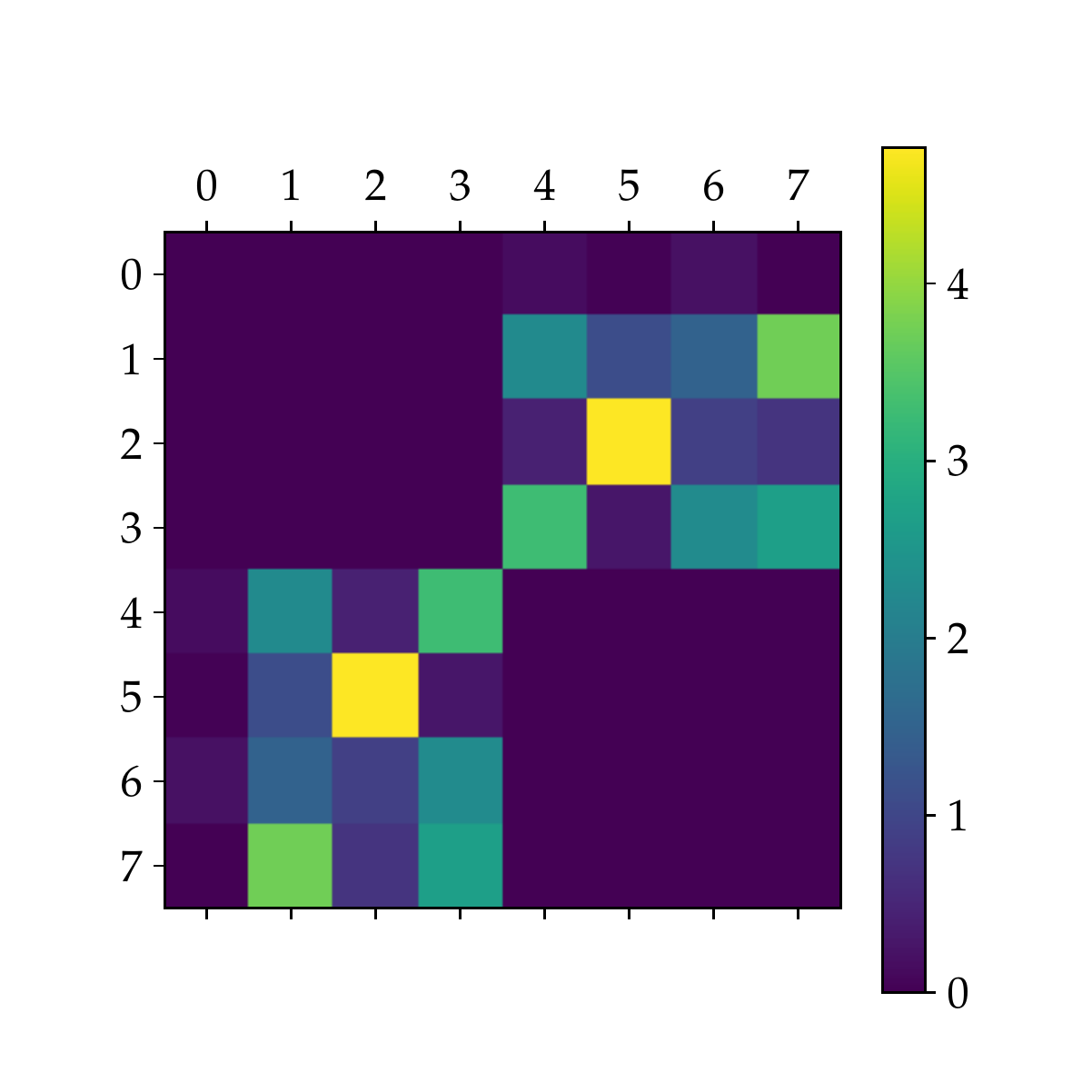}
\par\end{centering}
\caption{\label{fig:LogDipolarMatrix}The values of the dipolar matrix elements
$\log_{10} d_{i,j}$ obtained from the ground state of Si crystal (space group
227) via calculations using the LDA functional. }
\end{figure}

\begin{figure}
\begin{centering}
\includegraphics[width=8cm]{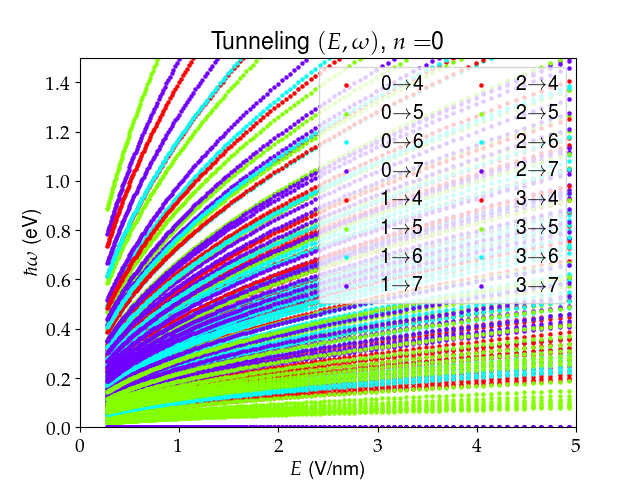}
\par\end{centering}
\caption{\label{fig:Selective-transition-removal}Selective transition removal
as a function of the laser parameters $\left(E,\hbar\omega\right)$
{[}$E$: field amplitude, $\hbar\omega$ is the photon energy{]}
for the transitions of zero-th order ($n=0$). }
\end{figure}

In a two-band description as in the Keldysh theory, suppression of the transition from one
band to the other may take place when the Rabi frequency is twice the
laser photon frequency \citep{Tamaya2019,Kato2020,Xia2020}
\begin{equation}
J_{n}\left(\frac{2\Omega_{\text{Rabi}}}{\omega_{\text{laser}}}\right)=0.\label{eq:Bessel1}
\end{equation}
However, when several electronic energy levels are available, a simultaneous
suppression of all possible transitions (a transition is denoted by
its dipolar matrix element $d_{i\rightarrow j}$) may not be feasible.
Assuming that the total transition probability can be reasonably
described by a series of Bessel functions in the physical reality, the
\emph{Le Bourget} theorem \citep{Watson1995}
suggests that \emph{only one} transition could be possibly disabled
for a given set of laser parameters (electric field of the wave, wavelength).
This mathematical argument provides a physical explanation for rather smooth reduction of total transition probability $w_{\text{PI}}$
observed in multiband description such as TDDFT (Fig. 3(b)) while
a two-band description such as the Keldysh model reveals a clear suppression
of transitions at given resonant frequencies {[}Fig. 3(a){]}. Then, since the Rabi frequency is proportional to $\Omega_{\text{Rabi}}\propto E\cdot d_{i,j}$
($E$ is the field amplitude), Eq. (\ref{eq:Bessel1}) can be rewritten
as 
\begin{equation}
J_{n}\left(\frac{2E\cdot d_{i,j}}{\hbar\omega^{2}}\right)=0\label{eq:BesselFunctionZeros}
\end{equation}
where $J_{n}\left(x\right)$ denotes the Bessel function of $n$-th
order. The dipolar transition matrix elements $d_{i,j}=\left\langle \psi_{i}|\hat{r}|\psi_{j}\right\rangle $
were obtained from the DFT computation. Figure \ref{fig:LogDipolarMatrix}
depicts the value of the dipolar matrix elements for Si in atomic
units. The elements indexed from 0 to 3 correspond to the valence
band states, the rest correspond to the conduction states. Since the description
of the Hamiltonian is hermitian, the matrix elements are symmetric above and below the matrix diagonal,
evidencing the equivalent transition probabilities for excitation and
recombination in the linear regime. 
Knowing the nodes of the Bessel functions of $n$-th order,
an illustrative mapping of the transition to be selectively disabled
can be provided (Fig. \ref{fig:Selective-transition-removal}). 

\section{Simplified model of laser energy absorption\label{sec:Drude-model}}

\paragraph{Energy absorbed by electrons.}

The energy $\xi_{\text{el}}^{\text{TDDFT}}\left(t\right)$
absorbed by the electrons is calculated as the difference between time-dependent electron
energy $e_{\text{el}}^{\text{TDDFT}}\left(t\right)$ and the ground
state energy $\xi_{\text{el}}^{\text{TDDFT}}\left(t\right)=e_{\text{el}}^{\text{TDDFT}}\left(t\right)-e_{\text{el}}^{\text{TDDFT}}\left(t=0\right)$.
To calculate the total energy of electrons $e_{\text{el}}^{\text{TDDFT}}\left(t\right)$, the formula given in the Appendix H of Ref. \cite{Martin2004} was used (it is also given in Chapter 5 of Ref. \cite{Ullrich2012}).

\paragraph{Electron excess energy.}

The energy density $de\left(\boldsymbol{r},t\right)$ introduced into
the sample by the laser at each time interval $dt$ can be expressed
using the total current density $\boldsymbol{j}\left(\boldsymbol{r},t\right)$
and the laser electric field $\boldsymbol{E}\left(\boldsymbol{r},t\right)$ as

\begin{align}
\frac{de\left(\boldsymbol{r},t\right)}{dt} & =\underbrace{\boldsymbol{j}\left(\boldsymbol{r},t\right)\cdot\boldsymbol{E}\left(\boldsymbol{r},t\right)}_{\text{total absorption (inter+intra)}}.\label{eq:TotalPowerDensity-1}
\end{align}
Usually the total current is separated to contributions from the
polarization (originating from interband transitions) and the free
carrier current $\boldsymbol{j}_{\text{free }}\left(t\right)$ related
to intraband processes. The dynamics of the absorbed energy density $e\left(\boldsymbol{r},t\right)$
can be described as
\begin{align}
\frac{de\left(\boldsymbol{r},t\right)}{dt} & \simeq\underbrace{n\hbar\omega\times w_{\text{PI}}\left(\boldsymbol{r},t\right)}_{\text{interband absorption}}+\label{eq:TotalPowerDensity}\\
 & +\underbrace{\boldsymbol{j}_{\text{free}}\left(\boldsymbol{r},t\right)\cdot\boldsymbol{E}\left(\boldsymbol{r},t\right)}_{\text{intraband absorption}},\nonumber 
\end{align}
where $w_{\text{PI}}\left(\boldsymbol{r},t\right)$ is the instantaneous
photoionization rate and $n=\left\lceil \frac{E_{\text{gap}}}{\hbar\omega}\right\rceil $
is the number of photons that are required to overcome the bare band
gap energy $E_{\text{gap}}$ using the electric field of the laser wave at frequency
$\omega$. In order to compare with the total energy, calculated by solving
the KS equation (Eq. \ref{eq:KSequation}), we integrate Eq. (\ref{eq:TotalPowerDensity})
over time in a volume $V$ as 
\begin{gather}
\xi_{\text{el}}\left(t\right)=\int_{V}d^{3}r\,e\left(\boldsymbol{r},t\right)\simeq\int_{V}d^{3}r\,n\hbar\omega\times n_{\text{exc}}\left(\boldsymbol{r},t\right)+\label{eq:TotalEnergy}\\
+\int_{V}d^{3}r\int_{0}^{t}dt'\,\boldsymbol{j}_{\text{free}}\left(\boldsymbol{r},t'\right)\cdot\boldsymbol{E}\left(\boldsymbol{r},t'\right).\nonumber 
\end{gather}
The density $n_{\text{exc}}$ is calculated by Eq. (\ref{eq:ExcitedElDensityDefinition}). Using the dipolar approximation for the external field, we obtain
the energy $\xi_{\text{el}}$ received by electrons from the laser field, which is defined as
\begin{gather}
\xi_{\text{el}}\left(t\right)=\int_{V}d^{3}r\,n\hbar\omega\times n_{\text{exc}}\left(\boldsymbol{r},t\right)+\label{eq:EnergyTime}\\
+\int_{0}^{t}dt'\,\boldsymbol{J}_{\text{free}}\left(t'\right)\cdot\boldsymbol{E}\left(t'\right).\nonumber 
\end{gather}

\paragraph{Time-dependent description of the free-carrier absorption}

The energy transferred from the laser light to the electrons via
intraband absorption can be computed from the excited electron density
$n_{\text{exc}}\left(t\right)$ and the conduction band electron current
$\boldsymbol{J}_{\text{free}}\left(t\right)$ by using of the generalized
Ohm law, expressed by
\[
\boldsymbol{J}_{\text{free}}\left(t\right)=\int dt'\sigma_{\text{free}}\left(t,t'\right)\boldsymbol{E}\left(t'\right).
\]
The contribution of the free electron currents can be described using a time-dependent conductivity $\sigma_{\text{free}}\left(\omega;t\right)=-i\omega\varepsilon_{0}\left[\varepsilon_{\text{Drude}}\left(\omega;t\right)-1\right]$,
where the dielectric permittivity $\varepsilon_{\text{Drude}}\left(t,\omega\right)$
is given by the Drude model for solids \cite{Stuart1996a} 
\begin{equation}
\varepsilon_{\text{Drude}}\left[n_{\text{exc}}\left(t\right)\right]\left(\omega\right)=1-\frac{e^{2}}{m_{e}m_{\text{eff}}\varepsilon_{0}\omega^{2}}\times\frac{n_{\text{exc}}\left(t\right)}{1+i\frac{\nu}{\omega}}.\label{eq:EpsilonDrude}
\end{equation}
The electron collision (or damping) time $\tau_D = \nu^{-1}$ is usually considered to be from one to several femtoseconds. In this work, the
effective electron mass is taken as given by the DFT [\citealp{LaflammeJanssen2016}], $m_{\text{eff}}=0.2226$. The collision time $\tau_D$ = 6 fs was adjusted to obtain the best fit to the TDDFT simulation results (see Fig. 4 of the main manuscript). 

\paragraph{Modeling the excess energy.}

Finally, with the above assumptions,  we derive from Eq. (\ref{eq:EnergyTime})
the balance equation for the excess electron energy $\xi_{\text{Drude}}$
based on the Drude model, which reads as
\begin{equation}
\frac{\partial\xi_{\text{Drude}}}{\partial t}=V\left[n\hbar\omega w_{\text{PI}}\left(t\right)+2\pi\varepsilon_{0}\int d\omega\,\omega\text{Im}\left[\varepsilon\left(\omega\right)\right]\boldsymbol{E}\left(\omega\right)^{2}\right].\label{eq:Drude}
\end{equation}
In the case of a quasi-continuous wave of frequency centered at $\omega_{0}$,
one has $\boldsymbol{E}\left(\omega\right)\rightarrow\delta\left(\omega\pm\omega_{0}\right)\boldsymbol{E}\left(\omega\right)$ that yields
\[
\frac{\partial\xi_{\text{Drude}}}{\partial t}=V\left[n\hbar\omega w_{\text{PI}}\left(t\right)+\varepsilon_{0}\omega_{0}\text{Im}\left[\varepsilon\left(\omega_{0}\right)\right]\boldsymbol{E}\left(\omega_{0}\right)^{2}\right].
\]
To calculate the absorbed electron energy at a time moment $t$ in a
volume $V$, we integrate the above expression over time from $t=0$ to $t$ that gives
\begin{gather}
\xi_{\text{Drude}}\left(t\right)=\nonumber \\
V\int_{0}^{t}dt'\left[n\hbar\omega w_{\text{PI}}\left(t'\right)+\varepsilon_{0}\omega_{0}\text{Im}\left\{ \varepsilon\left[n_{\text{exc}}\left(t'\right)\right]\left(\omega_{0}\right)\right\} \boldsymbol{E}\left(\omega_{0}\right)^{2}\right]\nonumber \\
=Vn\hbar\omega n_{\text{exc}}\left(t\right)+\label{eq:TotalEnergyAnalytical}\\
+V\int_{0}^{t}dt'\varepsilon_{0}\omega_{0}\text{Im}\left\{ \varepsilon\left[n_{\text{exc}}\left(t'\right)\right]\left(\omega_{0}\right)\right\} \boldsymbol{E}\left(\omega_{0}\right)^{2}.\nonumber 
\end{gather}
The later expression depends on the excited electron density $n_{\text{exc}}\left(t\right)$
and on three free parameters, namely the electron collision frequency $\nu$,
the electron effective mass $m_{\text{eff}}$, and the band gap energy
$E_{\text{gap}}$. We used Eq. (\ref{eq:TotalEnergyAnalytical})
to compare it with the electron excess energy $\xi_{\text{el}}^{\text{TDDFT}}$
obtained in our TDDFT simulations. 

\section{Excitation probabilities as a function of laser parameters\label{sec:Database-of-excitation}}

From the calculated dependences $n_{\text{exc}}^{\text{TDDFT}}\left(I_{\text{peak}}\right)$, the effective multiphoton rates~$\sigma_{n}$
associated with $n-$photon transitions can be extrapolated using the multiphoton approximation expressed by
\begin{equation}
\frac{\partial n_{\text{exc}}^{\text{TDDFT}}}{\partial t}=\sigma_{n}\frac{I^{n}}{n\hbar\omega}.\label{eq:NexcMacro}
\end{equation}
We remind that the LDA functional can underestimate the bandgap energy of crystals, which is usually smaller as compared to
more sophisticated modeling approaches for band structure calculations \citep{Waroquiers2013}.
Therefore, we have to note that here the effective multiphoton excitation rates $\sigma_{n}$ for silicon were calculated for the direct bandgap energy, which is somewhat smaller than the experimentally measured value. Using Eq. (\ref{eq:NexcMacro}),
one can fit the TDDFT results from Figs. 1(a), \ref{fig:Ibid-for-1600}(b), and \ref{fig:Ibid-for-800}(b), thus deriving the $\sigma_{n}$ values.
\begin{table*}
\begin{centering}
\begin{tabular}{cc|c|cc|cc}
 & \multicolumn{1}{c}{} & \multicolumn{1}{c}{} &  & \multicolumn{1}{c}{} &  & \tabularnewline
\hline 
\hline 
{\scriptsize{}Method} & {\scriptsize{}Wavelength} & {\scriptsize{}$\tau_p$} & {\scriptsize{}Band gap} & {\scriptsize{}Intensity range (W/cm$^{2}$)} & {\scriptsize{}Eff. transition probability} & {\scriptsize{}Ref.}\tabularnewline
\hline 
{\scriptsize{}Theory (TD-LDA)} & {\scriptsize{}3200 nm} & {\scriptsize{}30 fs} & {\scriptsize{}2.56 eV (d)} & {\scriptsize{}$\left(2.1-9.9\right)\times10^{10}$} & {\scriptsize{}$\sigma_{5}\left(\text{m}^{7}\text{W}^{-4}\right)=4.84\times10^{-56}$} & {\scriptsize{}This work}\tabularnewline
 &  &  &  & {\scriptsize{}$\left(1.0-2.6\right)\times10^{11}$} & {\scriptsize{}$\sigma_{4}\left(\text{m}^{5}\text{W}^{-3}\right)=3.05\times10^{-41}$} & {\scriptsize{}This work}\tabularnewline
 &  &  &  & {\scriptsize{}$\left(2.6-5.3\right)\times10^{11}$} & {\scriptsize{}$\sigma_{3}\left(\text{m}^{3}\text{W}^{-2}\right)=5.25\times10^{-26}$} & {\scriptsize{}This work}\tabularnewline
 &  &  &  & {\scriptsize{}$\left(0.53-3.4\right)\times10^{12}$} & {\scriptsize{}$\sigma_{2}\left(\text{m}\text{W}^{-1}\right)=2\times10^{-10}$} & {\scriptsize{}This work}\tabularnewline
 &  &  &  & {\scriptsize{}$\left(0.34-1.0\right)\times10^{13}$} & {\scriptsize{}$\sigma_{1}\left(\text{m}^{-1}\right)=2.93\times10^{6}$} & {\scriptsize{}This work}\tabularnewline
\cline{3-7} \cline{4-7} \cline{5-7} \cline{6-7} \cline{7-7} 
 &  & {\scriptsize{}20 fs} & {\scriptsize{}2.56 eV (d)} & {\scriptsize{}$\left(0.2-1.0\right)\times10^{11}$} & {\scriptsize{}$\sigma_{5}\left(\text{m}^{7}\text{W}^{-4}\right)=5.38\times10^{-56}$} & {\scriptsize{}This work}\tabularnewline
 &  &  &  & {\scriptsize{}$\left(1.0-2.6\right)\times10^{11}$} & {\scriptsize{}$\sigma_{4}\left(\text{m}^{5}\text{W}^{-3}\right)=3.42\times10^{-41}$} & {\scriptsize{}This work}\tabularnewline
 &  &  &  & {\scriptsize{}$\left(0.26-0.64\right)\times10^{12}$} & {\scriptsize{}$\sigma_{3}\left(\text{m}^{3}\text{W}^{-2}\right)=6.12\times10^{-26}$} & {\scriptsize{}This work}\tabularnewline
 &  &  &  & {\scriptsize{}$\left(0.6-3.4\right)\times10^{12}$} & {\scriptsize{}$\sigma_{2}\left(\text{m}\text{W}^{-1}\right)=2.52\times10^{-10}$} & {\scriptsize{}This work}\tabularnewline
 &  &  &  & {\scriptsize{}$\left(0.34-1\right)\times10^{13}$} & {\scriptsize{}$\sigma_{1}\left(\text{m}^{-1}\right)=3.94\times10^{6}$} & {\scriptsize{}This work}\tabularnewline
\cline{3-7} \cline{4-7} \cline{5-7} \cline{6-7} \cline{7-7} 
 &  & {\scriptsize{}10 fs} & {\scriptsize{}2.56 eV (d)} & {\scriptsize{}$\left(0.21-1.07\right)\times10^{11}$} & {\scriptsize{}$\sigma_{5}\left(\text{m}^{7}\text{W}^{-4}\right)=6.52\times10^{-56}$} & {\scriptsize{}This work}\tabularnewline
 &  &  &  & {\scriptsize{}$\left(1.07-2.9\right)\times10^{11}$} & {\scriptsize{}$\sigma_{4}\left(\text{m}^{5}\text{W}^{-3}\right)=4.37\times10^{-41}$} & {\scriptsize{}This work}\tabularnewline
 &  &  &  & {\scriptsize{}$\left(0.29-0.82\right)\times10^{12}$} & {\scriptsize{}$\sigma_{3}\left(\text{m}^{3}\text{W}^{-2}\right)=7.27\times10^{-26}$} & {\scriptsize{}This work}\tabularnewline
 &  &  &  & {\scriptsize{}$\left(0.82-3.4\right)\times10^{12}$} & {\scriptsize{}$\sigma_{2}\left(\text{m}\text{W}^{-1}\right)=3.92\times10^{-10}$} & {\scriptsize{}This work}\tabularnewline
 &  &  &  & {\scriptsize{}$\left(0.34-1\right)\times10^{13}$} & {\scriptsize{}$\sigma_{1}\left(\text{m}^{-1}\right)=6.38\times10^{6}$} & {\scriptsize{}This work}\tabularnewline
\cline{3-7} \cline{4-7} \cline{5-7} \cline{6-7} \cline{7-7} 
{\scriptsize{}Exp.} &  & {\scriptsize{}200 fs} &  &  & {\scriptsize{}$\sigma_{3}\left(\text{m}^{3}\text{W}^{-2}\right)=0.5\times10^{-26}$} & {\scriptsize{}Pearl et al. \citep{Pearl2008}}\tabularnewline
\hline 
{\scriptsize{}Theory (TD-LDA)} & {\scriptsize{}1600 nm} & {\scriptsize{}30 fs} & {\scriptsize{}2.56 eV (d)} & {\scriptsize{}$\left(0.11-1.6\right)\times10^{11}$} & {\scriptsize{}$\sigma_{4}\left(\text{m}^{5}\text{W}^{-3}\right)=4.41\times10^{-40}$} & {\scriptsize{}This work}\tabularnewline
 &  &  &  & {\scriptsize{}$\left(0.14-3.2\right)\times10^{12}$} & {\scriptsize{}$\sigma_{2}\left(\text{m}\text{W}^{-1}\right)=3.39\times10^{-10}$} & {\scriptsize{}This work}\tabularnewline
 &  &  &  & {\scriptsize{}$\left(0.32-1.6\right)\times10^{13}$} & {\scriptsize{}$\sigma_{1}\left(\text{m}^{-1}\right)=4.14\times10^{6}$} & {\scriptsize{}This work}\tabularnewline
\cline{3-7} \cline{4-7} \cline{5-7} \cline{6-7} \cline{7-7} 
 &  & {\scriptsize{}20 fs} & {\scriptsize{}2.56 eV (d)} & {\scriptsize{}$\left(0.11-1.6\right)\times10^{11}$} & {\scriptsize{}$\sigma_{4}\left(\text{m}^{5}\text{W}^{-3}\right)=4.62\times10^{-40}$} & {\scriptsize{}This work}\tabularnewline
 &  &  &  & {\scriptsize{}$\left(0.16-3.2\right)\times10^{12}$} & {\scriptsize{}$\sigma_{2}\left(\text{m}\text{W}^{-1}\right)=4.43\times10^{-10}$} & {\scriptsize{}This work}\tabularnewline
 &  &  &  & {\scriptsize{}$\left(0.32-1.3\right)\times10^{13}$} & {\scriptsize{}$\sigma_{1}\left(\text{m}^{-1}\right)=5.48\times10^{6}$} & {\scriptsize{}This work}\tabularnewline
\cline{3-7} \cline{4-7} \cline{5-7} \cline{6-7} \cline{7-7} 
{\scriptsize{}Exp.} &  & {\scriptsize{}200 fs} & {\scriptsize{}1.12 eV (i)} &  & {\scriptsize{}$\sigma_{2}\left(\text{m}\text{W}^{-1}\right)=\text{1.9\ensuremath{\times10^{-11}}}$} & {\scriptsize{}Bristow et al. \citep{Bristow2007}}\tabularnewline
\hline 
{\scriptsize{}Theory (TD-LDA)} & {\scriptsize{}800 nm} & {\scriptsize{}30 fs} & {\scriptsize{}2.56 eV (d)} & {\scriptsize{}$\left(0.12-14\right)\times10^{10}$} & {\scriptsize{}$\sigma_{2}\left(\text{m}\text{W}^{-1}\right)=8.05\times10^{-9}$} & {\scriptsize{}This work}\tabularnewline
 &  &  &  & {\scriptsize{}$\left(0.01-1.4\right)\times10^{13}$} & {\scriptsize{}$\sigma_{1}\left(\text{m}^{-1}\right)=5.72\times10^{6}$} & {\scriptsize{}This work}\tabularnewline
\cline{3-7} \cline{4-7} \cline{5-7} \cline{6-7} \cline{7-7} 
 &  & {\scriptsize{}20 fs} & {\scriptsize{}2.56 eV (d)} & {\scriptsize{}$\left(0.12-14\right)\times10^{10}$} & {\scriptsize{}$\sigma_{2}\left(\text{m}\text{W}^{-1}\right)=8.99\times10^{-9}$} & {\scriptsize{}This work}\tabularnewline
 &  &  &  & {\scriptsize{}$\left(0.14-14\right)\times10^{12}$} & {\scriptsize{}$\sigma_{1}\left(\text{m}^{-1}\right)=8.02\times10^{6}$} & {\scriptsize{}This work}\tabularnewline
\cline{3-7} \cline{4-7} \cline{5-7} \cline{6-7} \cline{7-7} 
 &  & {\scriptsize{}10 fs} & {\scriptsize{}2.56 eV (d)} & {\scriptsize{}$\left(0.12-14\right)\times10^{10}$} & {\scriptsize{}$\sigma_{2}\left(\text{m}\text{W}^{-1}\right)=1.07\times10^{-8}$} & {\scriptsize{}This work}\tabularnewline
 &  &  &  & {\scriptsize{}$\left(0.14-14\right)\times10^{12}$} & {\scriptsize{}$\sigma_{1}\left(\text{m}^{-1}\right)=1.45\times10^{7}$} & {\scriptsize{}This work}\tabularnewline
\cline{3-7} \cline{4-7} \cline{5-7} \cline{6-7} \cline{7-7} 
{\scriptsize{}Exp.} & {\scriptsize{}550 nm to 620 nm} & {\scriptsize{}90 fs} & {\scriptsize{}1.12 eV (i)} &  & {\scriptsize{}$9\times10^{-11}<\sigma_{2}\left(\text{m}\text{W}^{-1}\right)<36\times10^{-11}$} & {\scriptsize{}Reitze et al. \citep{Reitze1990}}\tabularnewline
{\scriptsize{}Exp.} & {\scriptsize{}800 nm} & {\scriptsize{}200 fs} & {\scriptsize{}1.12 eV (i)} &  & {\scriptsize{}$\sigma_{2}\left(\text{m}\text{W}^{-1}\right)=1.9\times10^{-11}$} & {\scriptsize{}Bristow et al. \citep{Bristow2007}}\tabularnewline
{\scriptsize{}Exp.} & {\scriptsize{}800 nm} & {\scriptsize{}-} & {\scriptsize{}1.12 eV (i)} &  & {\scriptsize{}$\sigma_{2}\left(\text{m}\text{W}^{-1}\right)=6.8\times10^{-11}$} & {\scriptsize{}Sjodin et al. \citep{Sjodin1998}}\tabularnewline
{\scriptsize{}Exp.} & {\scriptsize{}1060 nm} &  & {\scriptsize{}1.12 eV (i)} &  & {\scriptsize{}$\sigma_{2}\left(\text{m}\text{W}^{-1}\right)=1.5\times10^{-11}$} & {\scriptsize{}T.F. Boggess \citep{Boggess1984}}\tabularnewline
\hline 
{\scriptsize{}Theory (TD-LDA)} & {\scriptsize{}483 nm} & {\scriptsize{}20 fs} & {\scriptsize{}2.56 eV (d)} & {\scriptsize{}$\left(0.01-100\right)\times10^{11}$} & {\scriptsize{}$\sigma_{1}\left(\text{m}^{-1}\right)=1.09\times10^{7}$} & {\scriptsize{}This work}\tabularnewline
{\scriptsize{}Theory (TD-TB09)} & {\scriptsize{}407.58 nm} & {\scriptsize{}20 fs} & {\scriptsize{}3.04 eV (d)} & {\scriptsize{}$\left(0.01-140\right)\times10^{11}$} & {\scriptsize{}$\sigma_{1}\left(\text{m}^{-1}\right)=1.074\times10^{7}$} & {\scriptsize{}This work}\tabularnewline
{\scriptsize{}Exp.} & {\scriptsize{}483 nm} & {\scriptsize{}-} & {\scriptsize{}1.12 eV (i)} &  & {\scriptsize{}$\sigma_{1}\left(\text{m}^{-1}\right)=2.16\times10^{6}$} & {\scriptsize{}E. D. Palik \citep{Palik1985}}\tabularnewline
{\scriptsize{}Exp.} & {\scriptsize{}407.58 nm} & {\scriptsize{}-} & {\scriptsize{}1.12 eV (i)} &  & {\scriptsize{}$\sigma_{1}\left(\text{m}^{-1}\right)=0.964\times10^{7}$} & {\scriptsize{}E. D. Palik \citep{Palik1985}}\tabularnewline
\hline 
\hline 
 & \multicolumn{1}{c}{} & \multicolumn{1}{c}{} &  & \multicolumn{1}{c}{} &  & \tabularnewline
\end{tabular}
\par\end{centering}
\caption{\label{tab:Two-photon-excitation-rates-LDA}Multiphoton excitation
rates ($\sigma_{1}$, ..., $\sigma_{5}$) derived from the results of the first-principles simulations for the corresponding intensity ranges. The results obtained with
LDA and TB09 functionals are reported. Note that the adjustments of
effective multiphoton coefficients were performed in regimes where
tunneling and saturation effects may play a role. Available experimental data are also reported.}
\end{table*}
The data are summarised in Table
\ref{tab:Two-photon-excitation-rates-LDA} for the intensities below the saturation regimes. Where possible, we provide comparison of our data with the multiphoton
excitation rates available in the literature. As one can see, a reasonable agreement is achieved although in experimental studies pure multiphoton excitation can be masked by other processes involved at longer pulse durations such as collisional ionization and phonon-assisted indirect transitions.

We notice that for 800 nm wavelength, the value of $\sigma_{2}$ obtained
using first-principle simulations somewhat decreases when increasing
the laser pulse duration and the same is observed for other wavelengths and other $\sigma_{n}$. This may originate from the dynamic behavior
of the excitation. Once states at the bottom of the conduction bands are populated, the excitation probability decreases rapidly, an effect known as the Burstein-Moss effect \citep{Shah1976}. Although
the multiphoton ionization rates $\sigma_{2}$ obtained in the first-principles calculations
show a dependence on pulse duration $\tau_p$, the $\sigma_{n}$ values mostly remain within the same order of magnitude. Also we have calculated the single- and two-photon absorption rates $\sigma_{1}$ and $\sigma_{2}$ using the LDA and TB09
functionals at wavelengths of 484 nm and 407.58 nm respectively, the latter in order to adjust the laser wavelength for the resonant excitation of single-photon transition, see Table I. 

In the tunneling regime where the photoionization rate obtained in the TDDFT simulations scales linearly with the peak intensity,
an effective tunneling ionization rate $\sigma_{\text{1}}$ (using
Eq. (\ref{eq:NexcMacro}), $n=1$) can be estimated as a function of pulse duration. For $\lambda=800$ nm, $\sigma_{1}=1.447\times10^{6}$
m$^{-1}$ at $\tau_p =10$ fs; $\sigma_{1}=7.985\times10^{6}$ $m^{-1}$ at $\tau_p =20$ fs;
$\sigma_{1}=5.754\times10^{6}$ m$^{-1}$ at $\tau_p =30$ fs (not reported in the Table).  

Note that the direct comparison of the TDDFT photoionization rates with the
experimental values is not straightforward since the most experimental measurements
involve also indirect transitions ($\Gamma\rightarrow L$). Therefore, the comparison with
experimental measurements is only qualitative here. Note also that indirect
band gap transition rates can be computed using the TDDFT by employing
a localized electric field, as was very recently shown by Noda et al \cite{Noda2019}.

The full database of pulse-averaged photoionization rates obtained by our TDDFT simulations is
available at \url{http://www.quantumlap.eu/photo-ionization-database/}.

\bibliography{supplinf}